\documentclass[11pt]{article}
\usepackage{amssymb,a4wide}

\newcommand{\USU}{\mbox{$\mbox{U}[\mbox{sl}(2)]$}}
\newcommand{\UQSU}{\mbox{$\mbox{U}_{q}[\mbox{sl}(2)]$}}
\newcommand{\OP}[2]{\stackrel{\frown}{\prod_{#1}^{#2}}}
\newcommand{\BIN}[2]{\mbox{({\small
\raisebox{0.25ex}{$\begin{array}[c]{@{}c@{}}\scriptscriptstyle
#1\\[-1.5ex]\scriptscriptstyle 
#2\end{array}$}})}}
\newcommand{\DIM}{\BIN{N}{Q+N/2}}
\newcommand{\CC}{\mathbb{C}}
\newcommand{\ZZ}{\mathbb{Z}}
\newcommand{\EE}[3]{E^{#1}_{#2;#3}}
\newcommand{\Equer}[3]{\overline{E}^{#1}_{#2;#3}}
\newcommand{\E}[2]{\mathcal{E}^{#1}_{#2}}
\newcommand{\EH}[2]{\hat{\mathcal{E}}^{#1}_{#2}}
\newcommand{\PP}[3]{P^{#1}_{#2;#3}}
\newcommand{\Pquer}[3]{\overline{P}^{#1}_{#2;#3}}
\newcommand{\Fquer}[3]{\overline{F}^{#1}_{#2;#3}}
\newcommand{\F}[2]{\mathcal{F}^{#1}_{#2}}
\newcommand{\G}[2]{\mathcal{G}^{#1}_{#2}}
\newcommand{\GH}[2]{\hat{\mathcal{G}}^{#1}_{#2}}
\newcommand{\GT}[2]{\tilde{\mathcal{G}}^{#1}_{#2}}
\newcommand{\D}[2]{D^{#1}_{#2}}
\newcommand{\DDH}[2]{\hat{\mathcal{D}}^{#1}_{#2}}
\newcommand{\C}[2]{\chi_{#1,#2}}
\newcommand{\DEL}[2]{\Delta_{#1,#2}}
\newcommand{\OM}[2]{\Omega_{#1,#2}}
\newcommand{\zquer}{\overline{z}}
\newcommand{\zz}{(z,\zquer)}
\newcommand{\Nti}{\lim_{N\rightarrow\infty}}
\newcommand{\EPS}[3]{\epsilon^{#1}_{#2;#3}}
\newcommand{\EPSquer}[3]{\overline{\epsilon}^{#1}_{#2;#3}}
\newcommand{\ul}[1]{\underline{#1}}
\newcommand{\ol}[1]{\overline{#1}}
\newcommand{\sx}[1]{\sigma^{x}_{#1}}
\newcommand{\sy}[1]{\sigma^{y}_{#1}}
\newcommand{\sz}[1]{\sigma^{z}_{#1}}
\newcommand{\sP}[1]{\sigma^{+}_{#1}}
\newcommand{\sM}[1]{\sigma^{-}_{#1}}
\newcommand{\spm}[1]{\sigma^{\pm}_{#1}}
\newcommand{\Sz}{S^{z}_{}}
\newcommand{\Spm}{S^{\pm}_{}}
\newcommand{\ru}{\rule[0ex]{0ex}{4ex}}
\newcommand{\rd}{\rule[-2ex]{0ex}{2ex}}
\newcommand{\rud}{\rule[-2ex]{0ex}{6ex}}

\renewcommand{\theequation}{\arabic{section}.\arabic{equation}}

\begin{document}
\bibliographystyle{unsrt}

\begin{center}
\begin{Large}
\begin{bf}
The spin-1/2 XXZ Heisenberg chain,\\[0.25ex]
 the quantum algebra U$_{\mbox{{\boldmath $q$}}}$[sl(2)],\\[1ex] 
 and duality transformations for minimal models
\end{bf}
\end{Large}\vspace{6ex}

\begin{large}
{\bf Uwe Grimm}\\[2ex]
{\small Department of Mathematics,
University of Melbourne,\\
Parkville, Victoria 3052,
Australia}\\[4ex]
{\bf Gunter Sch\mbox{\"u}tz}\\[2ex]
{\small Department of Nuclear Physics,
  Weizmann Institute,\\
  Rehovot 76100,
  Israel}
\end{large}
\end{center}\vspace{4ex}

\begin{abstract}
The finite-size scaling spectra of the spin-$1/2$ XXZ Heisenberg chain
with toroidal boundary conditions and an even number of sites provide
a projection mechanism yielding the spectra of models with a central
charge $c<1$ including the unitary and non-unitary minimal series.
Taking into account the half-integer angular momentum sectors ---
which correspond to chains with an odd number of sites --- in many
cases leads to new spinor operators appearing in the projected
systems.  These new sectors in the XXZ chain correspond to a new type
of frustration lines in the projected minimal models.  The
corresponding new boundary conditions in the Hamiltonian limit are
investigated for the Ising model and the $3$-state Potts model and are
shown to be related to duality transformations which are an additional
symmetry at their self-dual critical point.  By different ways of
projecting systems we find models with the same central charge sharing
the same operator content and modular invariant partition function
which however differ in the distribution of operators into sectors and
hence in the physical meaning of the operators involved.

Related to the projection mechanism in the continuum there are
remarkable symmetry properties of the finite XXZ chain.  The observed
degeneracies in the energy and momentum spectra are shown to be the
consequence of intertwining relations involving \UQSU\ quantum algebra
transformations.
\end{abstract}\vfill

\centerline{\large This is a preprint version of 
J.~Stat.~Phys.~{\bf 71} (1993) 921--964}
\thispagestyle{empty}
\clearpage

\section{Introduction}
\setcounter{equation}{0}

Recently, the spin-$1/2$ XXZ Heisenberg chain has retained interest
since it was found \cite{toro} that the finite-size scaling limit
spectra of the chain with toroidal boundary conditions and an even
number of sites allow a projection mechanism that yields the spectra
of minimal unitary models with central charge $c<1$.  The projection
mechanism for the continuum models is based on the Feigin-Fuchs
construction \cite{FF} of the character functions of the Virasoro
algebra with central charge $c<1$ from the character functions with
$c=1$.  By taking differences of partition functions in the
finite-size scaling limit of the XXZ Hamiltonian with toroidal
boundary conditions (which corresponds to a free boson field theory
with central charge $c=1$), one obtains partition functions for models
with a central charge $c<1$.  This can be done in various ways,
yielding two classes of models which we call the $R$ and the $L$
models and which in turn are each divided into infinite series of
models labelled by a positive integer, also denoted by $R$ and $L$,
respectively.

But it has also been realized \cite{toro} that the projection
mechanism has a meaning for {\em finite}\/ systems as well. In many
cases, huge degeneracies in the spectra allow an analogous operation
on the finite-size spectra, where instead differences of partition
functions one considers differences of spectra (where this means
differences in terms of sets of eigenvalues).  This means that one
throws away all degenerate levels keeping only singlets. In this way,
one can obtain for instance the {\em exact}\/ finite-size spectra of
the Ising and $3$-states Potts quantum chains with $N$ sites from the
spectra of the XXZ chain with suitably chosen anisotropy and boundary
conditions with $2N$ sites (see \cite{toro}).

Although the spin-$1/2$ XXZ Heisenberg chain with toroidal boundary
conditions is not invariant under the quantum algebra \UQSU , it has
been realized \cite{PS} that part of the degeneracies related to the
projection mechanism which were observed for finite chains \cite{toro}
can be explained by investigation of the action of \UQSU\
transformations. In this paper, we show that this is indeed true for
{\em all}\/ the degeneracies observed in \cite{toro} and that in this
way one can explain the degeneracy of both the energy and the momentum
eigenvalues of the respective levels.  This answers part of the
questions left open in ref.~\cite{toro}.

Similar projection mechanisms as the one outlined above have been
found in the XXZ Heisenberg chain with free boundary conditions and
appropriately chosen surface fields at the ends of the chain
\cite{free1,free2,free3} (including a special choice with a \UQSU
-invariant Hamiltonian) and for a \mbox{spin-1} quantum chain
\cite{spin1} which allows to extract the spectra of systems belonging
to the minimal superconformal series.  Recently, similar structures
have been observed \cite{DKM} investigating the spectrum of the
$3$-states superintegrable chiral Potts model which is related to a
\mbox{spin-$1$} XXZ chain with anisotropy parameter
\mbox{$\gamma=\pi/3$}.

In this paper, we are going to generalize the projection mechanism of
ref.~\cite{toro}.  The class of systems obtained through the
projection procedure is enlarged to obtain models for all values $c<1$
of the central charge including the non-unitary minimal series.  Our
main emphasis, however, lies in the investigation of the same type of
projection mechanism, but now applied to half-integer angular momentum
sectors, i.e., we use the spectra of the XXZ chain with an odd number
of sites.  This results in the appearance of new spinor operators in
two classes of models and it therefore corresponds to new types of
frustration lines in the minimal models which are obtained by the
projection procedure. As explicit examples we study the Hamiltonian
limit of the Ising model and the $3$-state Potts model.  Here, we find
interesting new boundary conditions which have not been considered so
far and which turn out to be related to duality transformations.

Of special interest are also those models where the same operator
content can be obtained from an even and an odd number of sites. Our
numerical data for finite chains indicate that in these cases it is
not necessary to consider even and odd lengths in the finite chains
separately.  This observation agrees with results obtained for open
chains \cite{free3}.

Another result of our investigation is that we find systems which have
the same central charge, the same operator content and the same
modular invariant partition function but which differ in the
distribution of operators into sectors defined by the global
symmetries of the model.  This means that the physical significance of
these operators is different and therefore the operator content and
the modular invariant partition function alone are not sufficient to
completely characterize the universality class of a critical
system. One must take into account also the possible discrete
symmetries that are not determined by the partition function alone.

Our paper is organized as follows.  In sec.~2 we show how one can
obtain the spectra of $c<1$ systems by projection from the finite-size
scaling spectra of the XXZ Heisenberg chain in the continuum following
the scheme set up in ref.~\cite{toro}, our emphasis however lying on
the half-integer angular momentum sectors which in this context have
not been considered previously.  The $R$ and $L$ models are defined
and the operator content is given for the $R=1,2$ and $L=1,2$ models.
We illustrate the projection mechanism by means of explicit examples
for the $R=1$ and $L=1$ models where one obtains additional sectors
from the half-integer angular momentum sectors involving new spinor
operators in the projected systems.  Sec.~3 deals with the
implications of the projection mechanism for finite chains.  All the
degeneracies observed here and in ref.~\cite{toro} can be explained
using the representation theory of the quantum algebra \UQSU\ (see
appendix~A).  In sec.~4, we give an interpretation of the new sectors
obtained from half-integer spin sectors of the XXZ chain. Here, we
focus on the minimal models in the $R=1$ series.  As two explicit
examples we consider the Hamiltonian limit of the Ising model and of
the $3$-state Potts model. We show that for these cases one has to
choose a new type of boundary condition which is related to duality
transformations which in a sense act as a ``square root'' of the
translation operator.  Furthermore, we present some numerical data for
the $u=2$, $v=3$ model (central charge $c=-22/5$) which belongs to
that class of models for which the same operator content is obtained
from the scaling limit using an even and an odd number of sites in the
XXZ Heisenberg chain.

In the two appendices, we show that \UQSU\ quantum group
transformations explain the observed degeneracies between the spectra
of the finite XXZ Heisenberg chain with {\em different}\/
appropriately chosen toroidal boundary conditions.  For this purpose,
following the ideas of ref.~\cite{PS}, we establish intertwining
relations between different sectors of XXZ Hamiltonians with different
toroidal boundary conditions using the quantum algebra generators and
make use of the known structure of the irreducible representations of
\UQSU . Going beyond the results of \cite{PS}, this discussion also
includes the equality of the momenta of the levels concerned.
Furthermore, we give a short reminder of duality transformations for
the Ising and the 3-states Potts quantum chains.

\section{Projection mechanism in the finite-size scaling limit}
\setcounter{equation}{0}

Let us consider the spin-$1/2$ XXZ Heisenberg chain with a general
toroidal boundary condition ``$\alpha$'' defined by the Hamiltonian
\cite{ABB,oper}
\begin{equation} 
\begin{array}{rcr@{}l} 
{\displaystyle
H(q,\alpha,N)} & = & 
{\displaystyle -\,\frac{1}{2}\,} & {\displaystyle\left\{ \;
\sum_{j=1}^{N-1} \left( \sP{j}\sM{j+1} + \sM{j}\sP{j+1} +
(q+q^{-1})\sz{j}\sz{j+1} \right)\right. } \\ & & &
{\displaystyle\;\;\;\;\;\mbox{}\left.\left. 
\vphantom{\sum_{j=1}^{N-1}}
+ \;\alpha\,\sP{N}\sM{1}
+\,\alpha^{-1}\,\sM{N}\sP{1}
+\,(q+q^{-1})\,\sz{N}\sz{1}\;\right.\right\} }
\end{array}
\label{2e1} 
\end{equation} 
acting on a Hilbert space \mbox{${\cal H}(N) \cong (\CC^{2})^{\otimes
N}$}. Here, $N$ denotes the number of sites, $q$ and $\alpha$ are (for
the moment) arbitrary complex numbers and $\spm{j} = \sx{j} \pm
i\sy{j}$, where $\sx{j}$, $\sy{j}$, and $\sz{j}$ are the Pauli
matrices acting on the $j^{\mbox{th}}$ site of the chain.  Note that
the Hamiltonian $H(q,\alpha ,N)$ (which is related to the 6-vertex
model in the presence of a horizontal electric field
\cite{Y,SYY,McCW}) is hermitian if \mbox{$|\alpha | = 1$} and if
either $q$ is real or \mbox{$|q| = 1$}.

For arbitrary values of the parameters $q$ and $\alpha $, the
Hamiltonian (\ref{2e1}) commutes with the total spin (or ``charge'')
operator
\begin{equation}
\Sz\; =\; \frac{1}{2}\, \sum_{j=1}^{N} \sz{j} .
\label{2e3} 
\end{equation}
Therefore, we can split the Hilbert space ${\cal H}(N)$ into a direct
sum of $2N+1$ spaces with fixed value $Q$ of $\Sz$
\begin{equation}
{\cal H}(N)\; =\; \bigoplus_{Q=-N/2}^{Q=N/2} {\cal H}_{\, Q}(N)  ,
\label{2e4} 
\end{equation}
where $Q$ runs over the integer (half-integer) numbers depending on
$N$ being even (odd), respectively. We denote by ${\cal P}_{\, Q}$ the
projectors onto the subspaces ${\cal H}_{\, Q}$, $-N/2\leq Q\leq N/2$.
The dimension of ${\cal H}_{\, Q}(N)$ is given by \DIM , where
$\BIN{n}{k}=\frac{n!}{k!(n-k)!}$.

The Hamiltonian (\ref{2e1}) also commutes with a translation operator
\mbox{$T(\alpha ,N)$} which can be represented as an operator on
${\cal H}(N)$ in terms of Pauli matrices as follows
\begin{equation} 
T(\alpha, N) \; = \;
\alpha^{-\sz{1}/2} \cdot\OP{j=1}{N-1} P_{j} \; = \; \OP{j=1}{N-1}
P_{j} \;\cdot\;\alpha^{-\sz{N}/2} \; = \; \OP{j=1}{N-1}
\tilde{P}_{j}(\alpha, N) \;\cdot\;\alpha^{-\Sz /N}  .
\label{2e5} 
\end{equation}
Here, the operators $P_{j}$ (which permute the observables on sites
$j$ and $j+1$ in an obvious way) and $\tilde{P}_{j}$ are defined by
\begin{eqnarray}
P_{j} & = & \frac{1}{4} \left( \sP{j}\sM{j+1} + \sM{j}\sP{j+1}
        + 2 \left( \sz{j}\sz{j+1} + 1 \right) \right)
      \; =\; P_{j}^{\dagger}\; =\; P_{j}^{-1}, \label{2e6} \\[1ex]
\tilde{P}_{j}(\alpha ,N) & = &
\alpha^{-\sz{j}/2N}\; P_{j}\; \alpha^{\sz{j}/2N} 
\; = \; \alpha^{\sz{j+1}/2N}\; P_{j}\; \alpha^{-\sz{j+1}/2N} \nonumber \\
& = & \frac{1}{4} \left(
\alpha^{-1/N}\,\sP{j}\sM{j+1} + \alpha^{1/N}\,\sM{j}\sP{j+1}
      + 2 \left( \sz{j}\sz{j+1} + 1 \right) \right)  . \label{2e8}
\end{eqnarray}
where $j=1,\ldots,N-1$.  The symbol $\OP{\mbox{}}{\mbox{}}$ is used
for the ordered product
\begin{equation}
\OP{j=1}{N-1} P_{j}\; =\; P_{1} \cdot P_{2} \cdot \ldots \cdot P_{N-1}  .
\label{2e7} \end{equation}
The translation operator $T(\alpha ,N)$ is related to the momentum
operator $P(\alpha ,N)$ by
\begin{equation}
T(\alpha ,N)\; =\; \exp \left( -i \cdot P(\alpha ,N) \right)  .
\label{2e9} \end{equation}
Note that $T(\alpha ,N)$ is a unitary operator on ${\cal H}(N)$ (and
hence $P(\alpha ,N)$ is hermitian) if \mbox{$|\alpha | = 1$} and that
\mbox{$T(\alpha ,N)^{N} = \alpha^{-\Sz}$} and therefore is constant on
each subspace ${\cal H}_{\, Q}(N)$ (\ref{2e4}).

For later convenience we introduce a ``charge conjugation'' operator
$C$ defined by
\begin{equation}
C\; =\; \prod_{j=1}^{N} \sx{j}\; =\; 
C^{\dagger}\; =\; C^{-1} .
\label{2e10} 
\end{equation}
It satisfies the relations $C\cdot H(q,\alpha ,N)\cdot
C=H(q,\alpha^{-1},N)$, $C\cdot T(\alpha ,N)\cdot C=T(\alpha^{-1},N)$,
and $C\cdot\Sz\cdot C =-\Sz$.

We are now going to present an extended version of the projection
mechanism developed in ref.~\cite{toro}. For this purpose let us
recall the universal finite-size scaling limit partition function of
the spin-$1/2$ XXZ Heisenberg chain.  In what follows, we restrict
ourselves to $|q|=|\alpha |=1$ in eq.~(\ref{2e1}) which ensures the
hermiticity of both the Hamiltonian $H(q,\alpha ,N)$ (\ref{2e1}) and
the momentum operator $P(\alpha ,N)$ (\ref{2e9}).  We use the
parametrization
\begin{equation} 
q      \; = \; - \exp (-i\pi\gamma ),\qquad 
\alpha \; = \;   \exp (2\pi i\ell )
\label{3e1} \end{equation}
with two real numbers $0\leq\gamma <1$ and $-1/2<\ell\leq 1/2$.  The
Hamiltonian (\ref{2e1}) equipped with the normalization factor that
guarantees an isotropic continuum limit is given by
\begin{equation}
H^{\,\ell}(h,N)\; =\;\frac{\gamma}{2\sin (\pi\gamma )}\;
H(q=-\exp (-i\pi\gamma ), \alpha =\exp (2\pi i\ell ),N)  ,
\label{3e2} \end{equation}
where $h\geq 1/4$ is given by
\begin{equation}
h\; =\; \frac{1}{4} \left( 1-\gamma \right) ^{-1} .
\label{3e3} \end{equation}
The finite-size scaling limit of this system is known to be described
by the $c=1$ conformal field theory of a free compactified boson with
the compactification radius ${\cal R}$ being related to the anisotropy
$\gamma$ by \mbox{${\cal R}^{2}= 8 h$} (see e.g.\ ref.~\cite{BCR}).
Denoting the eigenvalues of $H^{\,\ell}_{\,
Q}(h,N)=H^{\,\ell}(h,N)\cdot {\cal P}_{\, Q}$ in the charge sector $Q$
with $-N/2\leq Q\leq N/2$ by $\EE{\ell}{Q}{j}(h,N)$ and the
corresponding momenta by $\PP{\ell}{Q}{j}(h,N)$, $j=1,2,\ldots ,\DIM
$, one obtains the following expression for the finite-size scaling
partition function of $H_{Q}^{\ell}(h,N)$ \cite{oper,HB}
\begin{equation} \begin{array}{rcl}
{\displaystyle \E{\ell}{Q}\zz}
& = & {\displaystyle\Nti \E{\ell}{Q}(z,\zquer ,N)}  \\
& = & {\displaystyle\Nti\sum_{j=1}^{\DIM}
z^{\,\frac{1}{2}(\Equer{\ell}{Q}{j}(h,N) + \Pquer{\ell}{Q}{j}(h,N))}\;
\zquer^{\,\frac{1}{2}(\Equer{\ell}{Q}{j}(h,N)
- \Pquer{\ell}{Q}{j}(h,N))}}   \\
& = & {\displaystyle\sum_{\nu\in\ZZ}\,
z^{\,\frac{[Q+4h(\ell +\nu )]^{2}}{16h}}\,\Pi_{V}(z)\;
\zquer^{\,\frac{[Q-4h(\ell +\nu )]^{2}}{16h}}\,\Pi_{V}(\zquer) .}
\end{array} \label{3e5} \end{equation}
In this equation, $\Equer{\ell}{Q}{j}(h,N)$ and
$\Pquer{\ell}{Q}{j}(h,N)$ denote the scaled gaps \cite{C}
\begin{eqnarray}
\Equer{\ell}{Q}{j}(h,N) & = & \frac{N}{2\pi}
\left( \EE{\ell}{Q}{j}(h,N) - E_{\, 0}(h,N) \right)  \label{3e6} \\
\Pquer{\ell}{Q}{j}(h,N) & = & \frac{N}{2\pi} \PP{\ell}{Q}{j}(h,N)
 ,  \label{3e7}
\end{eqnarray}
where $E_{\, 0}(h,N)=\EE{0}{0}{1}(h,N)$ is the ground-state energy of
the periodic Hamiltonianand $\Pi_{V}(z) = \prod_{n=1}^{\infty}
(1-z^{n}) ^{-1}$ is the generating function of the number of
partitions.  Eq.~(\ref{3e7}) has to be understood carefully. Since the
finite-size scaling partition function (\ref{3e5}) should only contain
the universal term of the partition function, one has to neglect
macroscopic momenta (i.e., momenta of order one).  To be more precise,
the levels that contribute to the partition function $\E{\ell}{Q}\zz$
in eq.~(\ref{3e5}) have a macroscopic momentum of modulus $0$ or $\pi$
depending on $\nu$ being even or odd, i.e., eq.~(\ref{3e7}) should be
modified as follows
\begin{equation}
\Pquer{\ell}{Q}{j}(h,N) \; = \; \frac{N}{2\pi}
\left( \PP{\ell}{Q}{j}(h,N)
\;-\; \pi \kappa^{\,\ell}_{\, Q\, ;\, j\,}(N) \right) 
\label{3e7a}
\end{equation}
where \mbox{$\kappa^{\,\ell}_{\, Q\, ;\, j\,}(N) \in \{ 0,1\}$}
depending on the value of the macroscopic momentum of the
corresponding level. Note that the scaled momenta are defined modulo
$N$ (since the momenta are defined modulo $2\pi$).

It is now our aim to extract the $c<1$ character function of
irreducible representations of the Virasoro algebra out of the
partition functions $\E{\ell}{Q}\zz$ (\ref{3e5}).  We parametrize the
central charge $c<1$ (we consider only real values of $c$) by a
positive real number $m$ by
\begin{equation}
c\; =\; 1\; -\; \frac{6}{m(m+1)}  .
\label{3e9} \end{equation}
Then one has to distinguish between the case that $m$ is a rational
number (corresponding to minimal models) and the other case that $m$
is irrational (corresponding to non-minimal models, i.e., the number
of irreducible highest weight representations is infinite in this
case).  For {\em irrational}\/ values of $m$, the character functions
$\C{r}{s}(z)=\mbox{tr}(z^{L_{0}})$ ($L_{0}$ generates (besides the
central element $c$) the Cartan subalgebra of the Virasoro algebra)
for a highest weight representation with highest weight $\DEL{r}{s}$
\begin{equation}
\DEL{r}{s}\; =\; \frac{[(m+1)r-ms]^{2}-1}{4m(m+1)}
\label{3e10} \end{equation}
with $r,s=1,2,\ldots$ are given by
\begin{equation}
\C{r}{s}(z)\; =\; \left( z^{\,\DEL{r}{s}} - z^{-\DEL{r}{s}}\right)\,
\Pi_{V}(z).
\label{3e11} \end{equation}
If however $m$ is {\em rational}, say $m=u/v$ with coprime positive
integers $u$ and $v$, the characters are \cite{RC}
\begin{eqnarray}
\C{r}{s}(z) & = & \OM{r}{s}(z)\; -\; \OM{r}{-s}(z) 
\label{3e12} \\
\OM{r}{s}(z) & = & \sum_{\nu\in\ZZ} z^{\frac{[2u(u+v)\nu
+(u+v)r-us]^{2}-v^{2}}{4u(u+v)}} \Pi_{V}(z)
\label{3e13} \end{eqnarray}
and we can restrict the possible values of $r$ and $s$ to the set
\begin{equation}
1 \;\leq\; r \;\leq\; u-1  ,\qquad
1 \;\leq\; s \;\leq\; u+v-1  .
\label{3e14}\end{equation}
The representations are unitary for integer values of $m$ \cite{FQS},
i.e., for $v=1$.

The shift of the central charge $c$ from the free boson value $c=1$ to
a value $c<1$ (\ref{3e9}) is now performed by choosing a new ground
state.  For this let us use the level $\EE{\ell_{0}}{0}{j_{0}}(h,N)$
in the charge sector $Q=0$ (note that we do not necessarily take the
lowest eigenvalue in this sector) of the Hamiltonian (\ref{3e2}) with
boundary condition $\ell_{0}$.  The number $j_{0}\geq 1$ has to be
chosen such that this level corresponds to the one that gives the
contribution $z^{\, h(\ell_{0}+\nu_{0})^{2}}\zquer^{\,
h(\ell_{0}+\nu_{0})^{2}}$ in the partition function (\ref{3e5}), where
$(\ell_{0}+\nu_{0})^{2}$ is related to $h$ by
\begin{equation}
c\; =\; 1\; -\; \frac{6}{m(m+1)}\; =\; 1\; -\; 24 h
(\ell_{0}+\nu_{0})^{2}
\label{3e15} \end{equation}
or equivalently to $\gamma$ by $\gamma
=1-m(m+1)(\ell_{0}+\nu_{0})^{2}$. (This means that $j_0$ labels the
level that contributes the difference of the universal parts of the
ground-state energies $\,
z^{\,\frac{1-c}{24}}\,\zquer^{\,\frac{1-c}{24}}\,$ between the
original XXZ Heisenberg chain with $c=1$ and the system we wish to
project out.)  Here, $-1/2\leq\ell_{0}\leq 1/2$ and $\nu_{0}$ is an
integer.  For definiteness, we choose the square root in
eq.~(\ref{3e15}) to be positive, hence
$\ell_{0}+\nu_{0}=(4hm(m+1))^{-1/2}$.

We now define {\em new}\/ scaled gaps with respect to our new ground
state $\EE{\ell_{0}}{0}{j_{0}}(h,N)$ with $h$ verifying
eq.~(\ref{3e15}) by (cf. eqs.~(\ref{3e6})--(\ref{3e7a}))
\begin{eqnarray}
\Fquer{k}{Q}{j}(N) & = & \frac{N}{2\pi} \left( \EE{k
(\ell_{0}+\nu_{0})}{Q}{j}(h,N) - \EE{\ell_{0}}{0}{j_{0}}(h,N) \right)
\label{3e17} \\ \Pquer{k}{Q}{j}(N) & = & \frac{N}{2\pi} \left( \PP{k
(\ell_{0}+\nu_{0})}{Q}{j}(h,N) - \pi \tilde{\kappa}^{\,k}_{\, Q\, ;\,
j\,}(N) \right) , \label{3e18}
\end{eqnarray}
where $k$ for the moment is arbitrary and
\mbox{$\tilde{\kappa}^{\,k}_{\, Q\, ;\, j\,} = \kappa^{\,k
(\ell_{0}+\nu_{0})}_{\, Q\, ;\, j\,} - \kappa^{\,\ell_{0}+\nu_{0}}_{\,
0\, ;\, j_{0}\,}$}.  Note that the momentum
$\PP{\ell_{0}}{0}{j_{0}}(h,N)$ of the new ground state always vanishes
(again this is only true up to a possible shift of $\pi$, see the
remark concerning eqs.~(\ref{3e7}) and (\ref{3e7a})).  The
corresponding finite-size scaling partition function is given by
(cf. eq.~(\ref{3e5}))
\begin{eqnarray}
\F{k}{Q}\zz & = & \Nti \F{k}{Q}(z,\zquer ,N) \nonumber \\ & = &
\Nti\sum_{j=1}^{\DIM} z^{\,\frac{1}{2}(\Fquer{k}{Q}{j}(N) +
\Pquer{k}{Q}{j}(N))}\; \zquer^{\,\frac{1}{2}(\Fquer{k}{Q}{j}(N) -
\Pquer{k}{Q}{j}(N))} \label{3e19} \\ & = & \sum_{\nu\in\ZZ}
z^{\frac{[m(m+1)(\ell_{0}+\nu_{0})Q+k+\frac{\nu}{\ell_{0}+\nu_{0}}]^{2}
-\, 1}{4m(m+1)}} \Pi_{V}(z)\;
\zquer^{\frac{[m(m+1)(\ell_{0}+\nu_{0})Q-k-\frac{\nu}{\ell_{0}+\nu_{0}}]^{2}
-\, 1}{4m(m+1)}} \Pi_{V}(\zquer) . \nonumber
\end{eqnarray}

We are now in a position to obtain the finite-size scaling partition
function of the unitary and non-unitary models with central charge
less than one. Eq.~(\ref{3e15}) gives $c$ as a function of the two
free (real) parameters $h$ and $\ell_{0} + \nu_{0}$.  Following
closely the results of ref.~\cite{toro}, we define two classes of
series of $c<1$ models by relating $h$ and $\ell_{0} + \nu_{0}$
through
\begin{equation}\label{3e19a}
\ell_0 + \nu_0\; =\; \frac{1}{M}\; -\; \frac{M}{4h}
\end{equation}
which we shall call the ``$R$''- models if $M>0$ and the
``$L$''-models if $M<0$. We label these two series of models by the
positive number $R=M$ or $L=-M$ resp. ($M\neq 0$) which will be
integer throughout this paper\footnote{Note that we change the
notation compared to ref.~\cite{toro} at this point.}.  We first
investigate the $R$-models ($M=R>0$).

\subsection{The \mbox{\protect\boldmath $R$}-models}

According to eq.~(\ref{3e19a}) we define the $R$-models by the
following relation between $h$ and $m$ which defines the central
charge according to eq.~(\ref{3e9})
\begin{equation}
h\; =\; \frac{R^{2}}{4}\cdot\frac{m+1}{m}
\label{3e20}\end{equation}
which means that $h>R^{2}/4$ for all $m$.  Using eqs.~(\ref{3e3}) and
(\ref{3e15}) one obtains
\begin{equation}
\gamma           \; = \; 1 - \frac{m}{R^{2}(m+1)},\qquad
\ell_{0}+\nu_{0} \; = \;\frac{1}{R(m+1)}, 
\label{3e22} \end{equation}
where we shall consider only integer values of $R$.

For $m$ {\em irrational}, the finite-size scaling partition functions
$\F{k}{Q}\zz$ (\ref{3e19}) are given by (see eq.~(\ref{3e10}))
\begin{equation}
\F{k}{Q}\zz\; =\;
\sum_{\nu\in\ZZ} z^{\,\DEL{\nu R+k}{k-Q/R}}\,\Pi_{V}(z)\;
\zquer^{\,\DEL{\nu R+k}{k+Q/R}}\,\Pi_{V}(\zquer ) .
\label{3e23}\end{equation}
One recognizes that for integer values of $R$ one can recover the
character functions (\ref{3e11}) through
\begin{equation}
\D{Rf}{R^{2}g}\zz\; = \;\F{Rf}{R^{2}g}\zz\; -\; \F{Rg}{R^{2}f}\zz\; =
\;\sum_{r=1}^{\infty} \C{Rr}{R\, (f-g)}(z)\,\C{Rr}{R\, (f+g)}(\zquer
),
\label{3e24}\end{equation}
where $f$ and $g$ are integers with $f>0$ and $-f<g<f$.  These
quantities are the finite-size scaling partition functions of the
non-minimal models with central charge given by (\ref{3e9}). The
integer numbers $R^2g$ and $Rf$ label the boundary conditions and
sectors of the projected systems according their internal global
symmetries, $g=0$ corresponding to periodic boundary conditions.  Note
that a vacuum representation \mbox{($\C{1}{1}(z)\C{1}{1}(\zquer )$)}
only occurs for the case $R=1$ (corresponding to the $2_{R}$ models of
ref.~\cite{toro}). The same is true for the modular invariant
partition function
\begin{equation}
{\cal A}_{\, m}\zz\; =\;
\sum_{r,s=1}^{\infty} \C{r}{s}(z)\,\C{r}{s}(\zquer )
\label{3e25}\end{equation}
which for $R=1$ can be written as the sum of the sectors $\D{f}{0}$ of
eq.~(\ref{3e24})
\begin{equation}
{\cal A}_{\, m}\zz\; =\;
\sum_{f=1}^{\infty} \D{f}{0}\zz
\label{3e26}\end{equation}
which is impossible for $R\neq 1$.

We now turn to the case $m$ {\em rational}, $m=u/v$ with coprime
positive integers $u$ and $v$. Eqs.~(\ref{3e20}) and (\ref{3e22}) take
the form
\begin{equation}
h                \; = \; \frac{R^{2}}{4}\cdot\frac{u+v}{u},\qquad 
\gamma           \; = \; 1 - \frac{u}{R^{2}(u+v)},\qquad 
\ell_{0}+\nu_{0} \; = \; \frac{v}{R(u+v)}.
\label{3e29}\end{equation}
The finite-size scaling partition functions $\F{k}{Q}\zz$ now have the
periodicity property \mbox{$\F{k}{Q}\zz=\F{k\pm n}{Q}\zz$} with
the integer $n$ being defined by
\begin{equation}
n\; =\; R(u+v)
\label{3e31}\end{equation}
and $\ell_{0}+\nu_{0}=v/n$.  From ref.~\cite{toro} we know that
instead of the partition functions $\F{k}{Q}\zz$ we should consider
\begin{eqnarray}
\G{k}{Q}\zz & = & \sum_{\mu\in\ZZ} \F{k}{Q+n\mu}\zz \nonumber \\ & = &
\sum_{\mu ,\nu\in\ZZ}
z^{\frac{[u(u+v)(\frac{Q}{n}+\mu)+kv+n\nu]^{2}-v^{2}}{4u(u+v)}}\,\Pi_{V}(z)\;
\zquer^{\frac{[u(u+v)(\frac{Q}{n}+\mu )-kv-n\nu
]^{2}-v^{2}}{4u(u+v)}}\, \Pi_{V}(\zquer ) \nonumber \\ & = &
\G{k}{Q\pm n}\zz\; =\; \G{k\pm n}{Q}\zz\; =\; \G{n-k}{n-Q}\zz
\label{3e32} 
\end{eqnarray}
that is we sum over all charge sectors modulo $n$ (\ref{3e31}).  If
$u(u+v)/n=u/R$ is integer (i.e., $u$ is a multiple of $R$), one can
rewrite this expression as follows
\begin{equation} \begin{array}{rcl}
{\displaystyle\G{k}{Q}\zz} & = &
{\displaystyle\sum_{w=0}^{\frac{2u}{R}-1}}
\begin{array}[t]{l}
{\displaystyle\left\{\sum_{\mu\in\ZZ} z^{\frac{[2u(u+v)\mu
+u\frac{Q}{R}+kv+(u+v)Rw]^{2}-v^{2}}{4u(u+v)}}\,\Pi_{V}(z) \right\} }
\\ {\displaystyle\left\{\sum_{\nu\in\ZZ} \zquer^{\frac{[2u(u+v)\nu
+u\frac{Q}{R}-kv-(u+v)Rw]^{2}-v^{2}}{4u(u+v)}}\, \Pi_{V}(\zquer
)\right\} } \end{array}  \\ & = &
{\displaystyle\sum_{w=0}^{\frac{2u}{R}-1}\;
\OM{Rw+k}{k-Q/R}(z)\,\OM{Rw+k}{k+Q/R}(\zquer ) } .
\end{array}\label{3e35}\end{equation}
Comparing this with eq.~(\ref{3e23}) one realizes that there are again
differences of these partition functions that can be written as
bilinear expressions in the characters (\ref{3e12})
\begin{eqnarray}
\D{Rf}{R^{2}g}\zz
\; =\; \D{R\, (u+v-f)}{R\, (u+v-Rg)}\zz
& = & \G{Rf}{R^{2}g}\zz - \G{Rg}{R^{2}f}\zz \nonumber \\
& = & \sum_{w=1}^{\frac{u}{R}-1}
\C{Rw}{R\, (f-g)}(z)\,\C{Rw}{R\, (f+g)}(\zquer )  ,
\label{3e36}\end{eqnarray}
where now $f$ and $g$ are integers that satisfy the inequalities
\mbox{$1\leq Rf\leq u+v-1$} and \mbox{$|Rg|\leq\min\{ Rf-1 ,
u+v-1-Rf\}$}.  Here, the quantities are the finite-size scaling
partition functions of the minimal models with central charge $c=1-6
v^{2}/(u(u+v))$ in the sector $Rf$ with boundary condition $R^2g$,
$g=0$ corresponding to periodic boundary conditions in the projected
system. The unitary series is given by $v=1$. Note that as discussed
above $c$ does not depend on the choice of $R$. As will be shown below
different values for $R$ lead to different systems with the same
central charge.

In addition to these sectors there are sectors that occur only for
certain rational values of $m$ and partly also include the
half-integer charge sectors of the Hamiltonian (\ref{3e2}).  In what
follows, we limit our discussion of these possibilities to the $R=1$
and $R=2$ models (corresponding to the $2_{R}$ and $1_{R}$ models of
ref.~\cite{toro}, respectively). Let us start exploiting the $R=1$
models.

\subsubsection{The \mbox{{\boldmath $R=1$}} models}

The unitary subset (i.e., $v=1$) of the $R=1$ series was discussed in
\cite{toro} where it had been called $2_R$-series. It represents the
$p$-states Potts models with $p=(2\cos{(\pi/(u+1))})^2$.  With $n=u+v$
(see eq.~(\ref{3e31})) for general $v$, eq.~(\ref{3e36}) just becomes
\begin{equation}
\D{f}{g}\zz \, =\,
\D{n-f}{n-g}\zz \, = \,
\G{f}{g}\zz - \G{g}{f}\zz \, = \,
\sum_{r=1}^{u-1} \C{r}{f-g}(z)\, \C{r}{f+g}(\zquer )
\label{3e37}\end{equation}
for all values of $u$ and
\begin{equation}
1\leq f\leq n-1 ,\qquad
 |g|\leq\min\left\{ f-1, n-1-f\right\}
\label{3e38}\end{equation}
These sectors (involving only integer charge sectors of the XXZ
Heisenberg chain) have the same structure as those obtained in
ref.~\cite{toro}.

By a closer inspection of eq.~(\ref{3e35}) one realizes that one has
the following possibilities to use half-integer values for $k$ and $Q$
in the partition function $\G{k}{Q}\zz$ (\ref{3e32}): For \ul{$u$
even}, $k$ has to be integer and $Q$ may be either integer or
half-integer, for \ul{$u+v$ even}, $k$ and $Q$ have both to be integer
or half-integer numbers, whereas for the case \ul{$v$ even} only $k$
may be half-integer valued and $Q$ has to be an integer. Defining
\mbox{$\tau =(1-(-1)^{2\tilde{g}})/4$}, one obtains as a
generalization of eq.~(\ref{3e37})
\begin{equation}
\D{\tilde{f}}{\tilde{g}}\zz\, =\,
\D{n-\tilde{f}}{n-\tilde{g}}\zz\, =\,
\G{\tilde{f}}{\tilde{g}}\zz\, -\,
\G{\tilde{g}+\tau n}{\tilde{f}+\tau n}\, =\,
\sum_{r=1}^{u-1} \C{r}{\tilde{f}-\tilde{g}+\tau n}(z)\,
\C{r}{\tilde{f}+\tilde{g}+\tau n}(\zquer )
\label{3e40}\end{equation}
where now $\tilde{f}$ and $\tilde{g}$ take the values specified above
and have to be chosen such that the conditions (\ref{3e14}) are
fulfilled.  To decide if in this way we really get {\em new}\/ sectors
(remember that one has to fulfill the conditions (\ref{3e14})), we
have a closer look to the sectors appearing in eq.~(\ref{3e40})
beginning with the case $u$ even.\vspace{2ex}

\noindent (a)\quad \ul{$u$ even:}\newline Here one can rewrite the
sectors of eq.~(\ref{3e40}) involving half-integer charges as
\begin{equation}
\D{g}{f-\frac{n}{2}}\zz \, =\,
\D{n-g}{\frac{n}{2}-f}\zz \, = \,
\G{g}{f-\frac{n}{2}}\zz -
\G{f}{g+\frac{n}{2}}\zz  \, = \,
\sum_{r=1}^{u-1} \C{u-r}{f-g}(z)\, \C{r}{f+g}(\zquer )
\label{3e41}\end{equation}
where now $f$ and $g$ are integers again. Clearly these sectors
fulfill the conditions (\ref{3e14}) if $f$ and $g$ satisfy
eq.~(\ref{3e38}) and one obtains {\em new}\/ sectors in this way as
long as $u$ is not equal to two. In this special case the sectors
non-trivially coincide and one obtains the same characters from an
even and an odd number of sites.  This will be reflected by special
features of the finite-size spectra (see the discussion in sec.~4.2
below) as it has already been observed for the free chain case (see
ref.~\cite{free3}). As an example let us consider the simplest model
of this kind which is the non-unitary $u=2$, $v=3$ model corresponding
to a central charge of $c=-22/5$.  The sectors of eqs.~(\ref{3e37})
and (\ref{3e41}) are
\begin{equation}
\begin{array}{r@{\;}c@{\;}l@{\;}c@{\;}l@{\;}c@{\;}l@{\;}c@{\;}l@{\;}c@{\;}l@{\;}c@{\;}l}
\D{1}{0} & = & \D{4}{0} & = & \G{1}{0} - \G{0}{1} & = & (0,0) & = &
\G{0}{3/2} - \G{4}{5/2} & = & \D{0}{7/2} & = & \D{0}{3/2}\\ \D{2}{0} &
= & \D{3}{0} & = & \G{2}{0} - \G{0}{2} & = &
(-\frac{1}{5},-\frac{1}{5}) & = & \G{0}{1/2} - \G{3}{5/2} & = &
\D{0}{9/2} & = & \D{0}{1/2} \\ \D{2}{1} & = & \D{3}{4} & = & \G{2}{1}
- \G{1}{2} & = & (0,-\frac{1}{5}) & = & \G{4}{1/2} - \G{3}{3/2} & = &
\D{1}{9/2} & = & \D{4}{1/2} \\ \D{3}{1} & = & \D{2}{4} & = & \G{3}{1}
- \G{1}{3} & = & (-\frac{1}{5},0) & = & \G{1}{1/2} - \G{3}{7/2} & = &
\D{4}{9/2} & = & \D{1}{1/2}
\end{array}\label{3e42}\end{equation}
where we used the notation $(\DEL{r}{s},\DEL{r^{\prime}}{s^{\prime}})$
for the product of characters
$\C{r}{s}(z)\C{r^{\prime}}{s^{\prime}}(\zquer )$ according to the
highest weights of the corresponding irreducible representations of
the Virasoro algebra. As an example for the other possibility $u\neq
2$ let us look to the unitary model $u=4$, $v=1$ with a central charge
$c=7/10$. Since $n=u+v=5$ as in the example above, the structure of
the sectors is exactly the same but one obtains
\begin{equation}\begin{array}{rclclcl}
\D{1}{0} & = & \D{4}{0} & = & \G{1}{0} - \G{0}{1}
& = & (0,0) + (\frac{7}{16},\frac{7}{16}) +
(\frac{3}{2},\frac{3}{2}) \\
\D{2}{0} & = & \D{3}{0} & = & \G{2}{0} - \G{0}{2}
& = & (\frac{1}{10},\frac{1}{10}) + (\frac{3}{80},\frac{3}{80}) +
(\frac{3}{5},\frac{3}{5}) \\
\D{2}{1} & = & \D{3}{4} & = & \G{2}{1} - \G{1}{2}
& = & (0,\frac{3}{5}) + (\frac{7}{16},\frac{3}{80}) +
(\frac{3}{2},\frac{1}{10}) \\
\D{3}{1} & = & \D{2}{4} & = & \G{3}{1} - \G{1}{3}
& = & (\frac{1}{10},\frac{3}{2}) + (\frac{3}{80},\frac{7}{16})
+ (\frac{3}{5},0)  \\
\D{0}{3/2} & = & \D{0}{7/2}
& = & \G{0}{3/2} - \G{4}{5/2}
& = & (\frac{3}{2},0) + (\frac{7}{16},\frac{7}{16}) +
(0,\frac{3}{2}) \\
\D{0}{1/2} & = & \D{0}{9/2}
& = & \G{0}{1/2} - \G{3}{5/2}
& = & (\frac{3}{5},\frac{1}{10}) + (\frac{3}{80},\frac{3}{80})
+ (\frac{1}{10},\frac{3}{5}) \\
\D{4}{1/2} & = & \D{1}{9/2}
& = & \G{4}{1/2} - \G{3}{3/2}
& = & (0,\frac{1}{10}) + (\frac{7}{16},\frac{3}{80})
+ (\frac{3}{2},\frac{3}{5}) \\
\D{1}{1/2} & = & \D{4}{9/2}
& = & \G{1}{1/2} - \G{3}{7/2}
& = & (\frac{1}{10},0) +(\frac{3}{80},\frac{7}{16}) +
(\frac{3}{5},\frac{3}{2})
\end{array}\label{3e43}\end{equation}

The half-integer charge sectors do not coincide with any integer
charge sector.\vspace{2ex}

\noindent (b)\quad\ul{$u+v$ even:}\newline Here we consider the case
that both $\tilde{f}$ and $\tilde{g}$ in eq.~(\ref{3e40}) are
half-integer numbers. One obtains
\begin{equation}
\D{\tilde{f}-\frac{n}{2}}{\tilde{g}}\zz\, =\,
\D{\frac{n}{2}-\tilde{f}}{n-\tilde{g}}\zz\, =\,
\G{\tilde{f}-\frac{n}{2}}{\tilde{g}}\, -\,
\G{\tilde{g}+\frac{n}{2}}{\tilde{f}}\, =\,
\sum_{r=1}^{u-1} \C{r}{\tilde{f}-\tilde{g}}(z)\,
\C{r}{\tilde{f}+\tilde{g}}(\zquer )
\label{3e44}\end{equation}
and therewith {\em new}\/ sectors if the half-integer numbers
$\tilde{f}$ and $\tilde{g}$ verify the same relations as the integers
$f$ and $g$ in eq.~(\ref{3e38}). Here, we take the $u=3$, $v=1$ model
which corresponds to the Ising model with central charge $c=1/2$ and
the $u=5$, $v=1$ model which corresponds to the $3$-state Potts model
with central charge $c=4/5$ (cf. ref.~\cite{toro}) as two examples.
From eqs.~(\ref{3e37}) and (\ref{3e44}) one obtains the following
sectors for $u=3$
\begin{equation}\begin{array}{rclclcl}
\D{1}{0} & = & \D{3}{0} & = & \G{1}{0} - \G{0}{1}
& = & (0,0) + (\frac{1}{2},\frac{1}{2}) \\
\D{2}{0} &   &           & = & \G{2}{0} - \G{0}{2}
& = & 2 \cdot (\frac{1}{16},\frac{1}{16}) \\
\D{2}{1} & = & \D{2}{3} & = & \G{2}{1} - \G{1}{2}
& = & (0,\frac{1}{2}) + (\frac{1}{2},0)  \\
\D{1/2}{1/2} & = & \D{7/2}{7/2}
& = & \G{1/2}{1/2} - \G{5/2}{5/2}
& = & (\frac{1}{16},0) + (\frac{1}{16},\frac{1}{2})\\
\D{7/2}{1/2} & = & \D{1/2}{7/2}
& = & \G{7/2}{1/2} - \G{5/2}{3/2}
& = & (0,\frac{1}{16}) + (\frac{1}{2},\frac{1}{16})
\end{array}\label{3e45a}\end{equation}
and for $u=5$ one has
\begin{equation}\begin{array}{rclclcl}
\D{1}{0} & = & \D{5}{0} & = & \G{1}{0} - \G{0}{1}
& = & (0,0) + (\frac{2}{5},\frac{2}{5}) + (\frac{7}{5},\frac{7}{5}) +
      (3,3) \\
\D{2}{0} & = & \D{4}{0} & = & \G{2}{0} - \G{0}{2}
& = & (\frac{1}{8},\frac{1}{8}) + (\frac{1}{40},\frac{1}{40}) +
      (\frac{21}{40},\frac{21}{40}) +
      (\frac{13}{8},\frac{13}{8}) \\
\D{3}{0} &  &           & = & \G{3}{0} - \G{0}{3}
& = & 2 \cdot (\frac{2}{3},\frac{2}{3}) +
      2 \cdot (\frac{1}{15},\frac{1}{15}) \\
\D{2}{1} & = & \D{4}{5} & = & \G{2}{1} - \G{1}{2}
& = &  (0,\frac{2}{3}) + (\frac{2}{5},\frac{1}{15}) +
       (\frac{7}{5},\frac{1}{15}) + (3,\frac{2}{3}) \\
\D{3}{1} & = & \D{3}{5} & = & \G{3}{1} - \G{1}{3}
& = &  (\frac{1}{8},\frac{13}{8}) +(\frac{1}{40},\frac{21}{40}) +
        (\frac{21}{40},\frac{1}{40}) +
       (\frac{13}{8},\frac{1}{8}) \\
\D{4}{1} & = & \D{2}{5} & = & \G{4}{1} - \G{1}{4}
& = &  (\frac{2}{3},0) + (\frac{1}{15},\frac{2}{5}) +
        (\frac{1}{15},\frac{7}{5}) + (\frac{2}{3},3) \\
\D{3}{2} & = & \D{3}{4} & = & \G{3}{2} - \G{2}{3}
& = &  (0,3) + (\frac{2}{5},\frac{7}{5}) + (\frac{7}{5},\frac{2}{5}) +
       (3,0)  \\
\D{1/2}{1/2} & = & \D{11/2}{11/2}
& = & \G{1/2}{1/2} - \G{7/2}{7/2}
& = &  (\frac{2}{3},\frac{1}{8}) + (\frac{1}{15},\frac{1}{40}) +
       (\frac{1}{15},\frac{21}{40}) + (\frac{2}{3},\frac{13}{8})\\
\D{3/2}{1/2} & = & \D{9/2}{11/2}
& = & \G{3/2}{1/2} - \G{7/2}{9/2}
& = & (\frac{1}{8},0) + (\frac{1}{40},\frac{2}{5}) +
      (\frac{21}{40},\frac{7}{5}) + (\frac{13}{8},3) \\
\D{9/2}{1/2} & = & \D{3/2}{11/2}
& = & \G{9/2}{1/2} - \G{7/2}{3/2}
& = &  (0,\frac{1}{8}) + (\frac{2}{5},\frac{1}{40}) +
       (\frac{7}{5},\frac{21}{40}) + (3,\frac{13}{8}) \\
\D{11/2}{1/2} & = & \D{1/2}{11/2}
& = & \G{11/2}{1/2} - \G{7/2}{5/2}
& = &  (\frac{1}{8},\frac{2}{3}) + (\frac{1}{40},\frac{1}{15}) +
      (\frac{21}{40},\frac{1}{15}) +
      (\frac{13}{8},\frac{2}{3}) \\
\D{1/2}{3/2} & = & \D{11/2}{9/2}
& = & \G{1/2}{3/2} - \G{9/2}{7/2}
& = &  (\frac{13}{8},0) + (\frac{21}{40},\frac{2}{5}) +
       (\frac{1}{40},\frac{7}{5}) + (\frac{1}{8},3) \\
\D{11/2}{3/2} & = & \D{1/2}{9/2}
& = & \G{11/2}{3/2} - \G{9/2}{5/2}
& = &  (0,\frac{13}{8}) + (\frac{2}{5},\frac{21}{40}) +
        (\frac{7}{5},\frac{1}{40}) + (3,\frac{1}{8})
\end{array}\label{3e45b}\end{equation}
Here, not only the half-integer charge sectors as such are new (i.e.,
the {\em combination}\/ of the various building blocks contributing to
it), in addition they are build by contributions from so far unknown
spinor fields with anomalous dimensions $(\Delta,\overline{\Delta})$
as given in (\ref{3e45a}) and (\ref{3e45b}).  The partition functions
of the form $\D{n/2}{0}$ are special in so far as the sector in the
XXZ Heisenberg chain from which they are obtained splits into two
subsectors with eigenvalue $C=\pm1$ of the charge conjugation operator
$C$ (\ref{2e10}). This symmetry is not used in the projection
mechanism and is presumably the reason why these partition functions
contain each contribution twice\footnote{We did not check this
assumption.}.  We will return to these models later when we study the
corresponding finite systems.\vspace{2ex}

\noindent (c)\quad\ul{$v$ even:}\newline In this case one would expect
new sectors appearing for half-integer values of $\tilde{f}$ in
eq.~(\ref{3e40}), but this is not the case due to the identity
\mbox{$\G{k\pm\frac{n}{2}}{Q}\zz\equiv\G{k}{Q}\zz$} which follows
simply from the fact that $\frac{n}{2}(\ell_{0}+\nu_{0})=\frac{v}{2}$
is an integer.

\subsubsection{The \mbox{{\boldmath $R=2$}} models}

Here, the unitary subset $v=1$ discussed in \cite{toro} (called $1_R$
models there) corresponds to the low-temperature $O(p)$-models
\cite{NRS,SALEUR} with $p=2\cos{(\pi/(u+1))}$.  We have to investigate
$u$ even and $u$ odd separately ($u$ odd was not discussed in
\cite{toro}) as we can use eq.~(\ref{3e35}) for $u$ {\em even}\/ only,
i.e., for $n=2(u+v)\equiv 2\bmod 4$ (since $v$ has to be odd).
Eq.~(\ref{3e36}) now reads
\begin{eqnarray}
\D{2f}{4g}\zz
\; =\; \D{n-2f}{n-4g}\zz
& = & \G{2f}{4g}\zz - \G{2g}{4f}\zz \nonumber \\
& = & \sum_{w=1}^{\frac{u}{2}-1}
\C{2w}{2(f-g)}(z)\,\C{2w}{2(f+g)}(\zquer )  .
\label{R2e1}\end{eqnarray}
There are however additional sectors besides these.  But only integer
charge sectors contribute here, since, although one can form bilinear
expressions in character functions from the partition functions of the
half-integer charge sectors, these involve negative multiplicities
which we do not want to consider in our present discussion.  For
completeness, we will briefly state the relevant
equations.\vspace{2ex}

\noindent (a)\quad\ul{$u$ even:}\newline On obtains the following
expressions for the $\G{k}{Q}\zz$ (\ref{3e35}) in terms of the
functions $\OM{r}{s}(z)$ (\ref{3e13}):
\begin{eqnarray}
\G{k}{Q}\zz & = & \;\sum_{d=1-\xi}^{\frac{u}{2}-\xi}
\OM{\xi +2d}{k-\frac{Q}{2}+\eta (u+v)}(z)\,
\OM{\xi +2d}{k+\frac{Q}{2}+\eta (u+v)}(\zquer ) \nonumber \\
& & + \sum_{d=0}^{\frac{u}{2}-1}
\OM{-(\xi +2d)}{k-\frac{Q}{2}+\eta (u+v)}(z)\,
\OM{-(\xi +2d)}{k+\frac{Q}{2}+\eta (u+v)}(\zquer )
\label{R2e2}\end{eqnarray}
where $\eta =(1-(-1)^{Q})/4$ and $\xi =(1-(-1)^{k})/2$ ($\xi
=(1-(-1)^{k+Q})/2$) if $u\equiv 0\bmod 4$ ($u\equiv 2\bmod 4$),
respectively.  From this, one gets in generalization of
eq.~(\ref{R2e1}) the following sectors
\begin{equation}
\D{2f+\omega}{4g+t}\zz \; = \; \G{2f+\omega}{4g+t}\zz -
\G{\frac{4g+t}{2}+
\varepsilon\frac{2\omega +t}{2}(u+v)}{2(2f+\omega )+t(u+v)}\zz
\label{R2e5} \end{equation}
with $\omega\in\{ 0,1\}$, $t\in\{ 0,1,2,3\}$, and $\varepsilon =
(-1)^{(u+v+1)/2}$.  In terms of character functions, they have the
form
\begin{equation}
\D{k}{Q}\zz\; = \;\sum_{d=1-\xi}^{\frac{u}{2}-1}
\C{\xi +2d}{k-\frac{Q}{2}+\eta (u+v)}(z)\,
\C{\xi +2d}{k+\frac{Q}{2}+\eta (u+v)}(\zquer )  . 
\label{R2e6}\end{equation}
For physical sectors, of course, the character functions which enter
on the right-hand side of eq.~(\ref{R2e6}) have to comply with the
conditions (\ref{3e14}), taking into account the periodicity
properties of the character functions $\C{r}{s} \zz$ in the indices
$r$ and $s$.\vspace{2ex}

\noindent (b)\quad\ul{$u$ odd:}\newline Although, as mentioned above,
eq.~(\ref{3e35}) does not apply for odd values of $u$ one can
construct $R=2$ models in this case, too. However, one has to consider
new partition functions $\GT{k}{Q}\zz$ which are the sums of two
sectors $\G{k}{Q}$
\begin{equation}
\GT{k}{Q}\zz\; = \;\G{k}{Q}\zz +
\G{k+\frac{n}{2}}{Q}\zz  ,
\label{R2e11}\end{equation}
in order to obtain suitable expressions.  Following the same procedure
which led to eq.~(\ref{3e35}), this yields
\begin{eqnarray}
\GT{k}{Q}\zz & = & \sum_{\mu ,\nu\in\ZZ}
\begin{array}[t]{l} {\displaystyle \left\{\;
z^{\frac{[u(u+v)\mu +\frac{u}{2}Q+kv+(u+v)\nu ]^{2}-v^{2}}{4u(u+v)}}\,
\Pi_{V}(z) \right. }  \\
{\displaystyle \left. \;\;
\zquer^{\frac{[u(u+v)\mu +\frac{u}{2}Q-kv-(u+v)\nu ]^{2}-v^{2}}{4u(u+v)}}\,
\Pi_{V}(\zquer) \;\right\} } \end{array} \nonumber \\
& = &
\sum_{w=0}^{u-1}\;
\OM{w+k}{k-Q/2}(z)\,\OM{w+k}{k+Q/2}(\zquer )  . 
\label{R2e13}\end{eqnarray}
Hence, the new partition functions $\GT{k}{Q}\zz$ are in fact the same
as the sectors $\G{k}{Q/2}\zz$ (\ref{3e35}) of the corresponding $R=1$
model with $n=u+v$. Of course all the equations obtained there
translate to the present case.

\subsection{The \mbox{\protect\boldmath $L$}-models}

Similar to the $R$-models we define the $L$-models by
\begin{equation}
h                \; = \; \frac{L^{2}}{4}\cdot\frac{m}{m+1},\qquad
\gamma           \; = \; 1 - \frac{m+1}{L^{2}m},\qquad
\ell_{0}+\nu_{0} \; = \; \frac{1}{Lm}, 
\label{3eL3}\end{equation}
where in what follows $L$ will be integer-valued again and $h<L^{2}/4$
for all possible values of $m$.

Let us commence by considering {\em irrational}\/ values of $m$.  The
finite-size scaling partition functions $\F{k}{Q}\zz$ (\ref{3e19}) in
this case are given by (\ref{3e10})
\begin{equation}
\F{k}{Q}\zz\; =\;
\sum_{\nu\in\ZZ} z^{\,\DEL{-k-Q/L}{\nu L-k}}\,\Pi_{V}(z)\,
\zquer^{\,\DEL{-k+Q/L}{\nu L-k}}\,\Pi_{V}(\zquer ) .
\label{3eL4}\end{equation}
The character functions (\ref{3e11}) are now recovered through
\begin{equation}
\D{Lf}{L^{2}g}\zz\;
=\; \F{Lf}{L^{2}g}\zz\; -\; \F{Lg}{L^{2}f}\zz\;
=\; \sum_{s=1}^{\infty} \C{L\, (f+g)}{Ls}(z)\,\C{L\, (f-g)}{Ls}(\zquer ),
\label{3eL5}\end{equation}
where again $f$ and $g$ are integers with $f>0$ and $-f<g<f$.  As in
the case of the $R$-models, the vacuum representation is included in
this set for $L=1$ only, and the modular invariant partition function
${\cal A}_{m}\zz$ (\ref{3e25}) for $L=1$ (in terms of the sectors
defined above) is given by the same expression (\ref{3e26}) as for the
$R=1$ models.

We now switch to {\em rational}\/ values of $m=u/v$ with coprime
positive integers $u$ and $v$ again. Eq.~(\ref{3eL3}) becomes
\begin{equation}
h                 \; = \;\frac{L^{2}}{4}\cdot\frac{u}{u+v}  ,\qquad
\gamma            \; = \; 1 - \frac{u+v}{L^{2}u}  ,\qquad
\ell_{0} +\nu_{0} \; = \;\frac{v}{Lu} \; = \; \frac{v}{n}  ,
\label{3eL9}\end{equation}
and the integer $n$ (cf. eq.~(\ref{3e31})) is given by $n=Lu$.  Once
again, we define partition functions $\G{k}{Q}$ in the same way as in
eq.~(\ref{3e32}) above. If $u(u+v)/n=(u+v)/L$ is an integer
(i.e., $u+v$ is a multiple of $L$), these can be rewritten as follows
(cf. eq.~(\ref{3e35}))
\begin{equation} 
\G{k}{Q}\zz\; = \;
\sum_{w=0}^{\frac{2(u+v)}{L}-1}\;
\OM{-k-Q/L}{Lw-k}(z)\,\OM{-k+Q/L}{Lw-k}(\zquer ) 
\label{3eL11}\end{equation}
and one obtains expressions bilinear in the characters (\ref{3e12}) by
\begin{eqnarray}
\D{Lf}{L^{2}g}\zz
\; =\; \D{L\, (u-f)}{L\, (u-Lg)}\zz
& = & \G{Lf}{L^{2}g}\zz - \G{Lg}{L^{2}f}\zz \nonumber \\
& = &  \sum_{w=1}^{\frac{u+v}{L}-1}
\C{L\, (f+g)}{Lw}(z)\,\C{L\, (f-g)}{Lw}(\zquer )  ,
\label{3eL12}\end{eqnarray}
where $f$ and $g$ are integers satisfying
\mbox{$1\leq Lf\leq u-1$} and 
\mbox{$|Lg| \leq \min\{ Lf-1 , u-1-Lf\}$}.

We now proceed by investigating the additional sectors that occur for
the $L=1$ and $L=2$ models (corresponding to the $2_{L}$ and $1_{L}$
models of ref.~\cite{toro}).  Note that since the range of $h$
(\ref{3e3}) in the XXZ chain is limited to $h\geq 1/4$, the $L=1$
models cannot be realized in the finite-size spectra of the XXZ
Heisenberg chain.

\subsubsection{The \mbox{{\boldmath $L=1$}} models}

The unitary subset ($v=1$) discussed in \cite{toro} corresponds to the
tricritical $p$-states Potts models, $p=(2\cos{(\pi/u}))^2$ ($2_L$
series). Again the structure of the sectors obtained in
ref.~\cite{toro} is identical to what one obtains from
eq.~(\ref{3eL12}) for $v$ not restricted to one but integer charge
sectors only.  Using $n=u$ for $L=1$, the sectors read
\begin{equation}
\D{f}{g}\zz\; =\;\D{n-f}{n-g}\zz\; =\;
\G{f}{g}\zz - \G{g}{f}\zz\; = \;
\sum_{s=1}^{u+v-1} \C{f+g}{s}(z)\, \C{f-g}{s}(\zquer )
\label{3eL13}\end{equation}
for any value of $u+v$ and
\begin{equation}
1\leq f\leq n-1 ,   |g|\leq\min\left\{ f-1,n-1-f\right\}
 .
\label{3eL14}\end{equation}

Again we want to include half-integer values for $k$ and $Q$.  Let
\mbox{$\tau =(1-(-1)^{2\tilde{g}})/4$} as in eq.~(\ref{3e40}).  One
obtains as a generalization of eq.~(\ref{3eL13})
\begin{equation}
\D{\tilde{f}}{\tilde{g}}\zz \; =\;
\D{n-\tilde{f}}{n-\tilde{g}}\zz \; =\;
\G{\tilde{f}}{\tilde{g}}\zz -
\G{\tilde{g}+\tau n}{\tilde{f}+\tau n}\zz \; =\;
\sum_{s=1}^{u+v-1} \C{r}{\tilde{f}-\tilde{g}+\tau n}(z)\,
\C{r}{\tilde{f}+\tilde{g}+\tau n}(\zquer ) \; .
\label{3eL15}\end{equation}
In this equation, $\tilde{f}$ is integer and $\tilde{g}$ integer or
half-integer if $u$ and $v$ are both even (i.e., for $u+v$ even); for
$u$ even and $v$ odd, $\tilde{f}$ and $\tilde{g}$ have to be both even
or odd whereas in the case of an even value of $v$, $\tilde{g}$ has to
be an integer ($\tilde{f}$ possibly being half-integer).  We again
consider these three cases separately.\vspace{2ex}

\noindent (a)\quad\ul{$u+v$ even:}\newline
Consider the sectors of eq.~(\ref{3eL15}) with half-integer charge.
They are
\begin{eqnarray}
\D{-g}{\frac{n}{2}-f}\zz \; =\;
\D{n+g}{\frac{n}{2}+f}\zz & = &
\G{-g}{\frac{n}{2}-f}\zz -
\G{-f}{\frac{n}{2}-g}\zz \nonumber \\ & = &
\sum_{s=1}^{u+v-1} \C{f+g}{u+v-s}(z)\, \C{f-g}{s}(\zquer )  ,
\label{3eL16}\end{eqnarray}
where $f$ and $g$ are integers which comply with eq.~(\ref{3eL14}).
Here, one always obtains {\em new}\/ sectors, since there is no
possibility to have $u+v=2$ which would be the analogue of the case
$u=2$ for the $R=1$ models (cf. eqs.~(\ref{3e41})--(\ref{3e42})).

As a simple example we consider the case $u=3$, $v=1$ with $n=3$ which
is a model with central charge $c=1/2$. One has
\begin{equation}\begin{array}{rclclcl}
\D{1}{0} & = & \D{2}{0} & = & \G{1}{0} - \G{0}{1}
& = & (0,0) + (\frac{1}{16},\frac{1}{16}) + (\frac{1}{2},\frac{1}{2})
 \\[1ex]
\D{0}{1/2} & = & \D{0}{5/2} & = & \G{0}{1/2} - \G{2}{3/2}
& = & (\frac{1}{2},0) + (\frac{1}{16},\frac{1}{16}) + (0,\frac{1}{2})
\end{array}\label{3eL17}\end{equation}
and one realizes that taking into account the half-integer charge
sectors one obtains the operator content of the Ising model but with a
different distribution of operators into sectors. For instance, the
leading thermal exponent is $1/8$ for this model whereas it is $1$ for
the Ising model which is realized as the corresponding $R=1$ model
(see eq.~(\ref{3e45a}) and ref.~\cite{toro}).\vspace{2ex}

\noindent (b)\quad\ul{$u$ even:}\newline The sectors with
half-integer values of $\tilde{f}$ and $\tilde{g}$ are given by
(cf. eq.~(\ref{3eL15}))
\begin{eqnarray}
\D{- \tilde{g}}{\frac{u}{2}- \tilde{f}}\zz \; =\;
\D{u+ \tilde{g}}{\frac{u}{2}+ \tilde{f}}\zz  & = &
\G{- \tilde{g}}{\frac{u}{2}- \tilde{f}}\zz -
\G{- \tilde{f}}{\frac{u}{2}- \tilde{g}}\zz \nonumber \\
& = &
\sum_{s=1}^{u+v-1} \C{\tilde{f}+\tilde{g}}{u+v-s}(z)\,
\C{\tilde{f}-\tilde{g}}{s}(\zquer )
\label{3eL18}\end{eqnarray}
which again are {\em new}\/ sectors compared to eq.~(\ref{3eL13})
provided the half-integer numbers $\tilde{f}$ and $\tilde{g}$ fulfill
the same relations as $f$ und $g$ in eq.~(\ref{3eL14}). There is one
exception: the models with $u=2$ ($n=2$) which due to
eq.~(\ref{3eL14}) consist of the sector $\D{1}{0}\zz$ alone.

To give an example for this class of models we consider the case
$u=2$, $v=3$ (hence $n=2$) which corresponds to a central charge
$c=-\frac{22}{5}$.  The only sector is
\begin{equation}\begin{array}{rclcl}
\D{1}{0} & = & \G{1}{0} - \G{0}{1}
& = & 2\cdot (0,0) + 2\cdot (-\frac{1}{5},-\frac{1}{5})  .
\end{array}\label{3eL19}\end{equation}
As an example for a model with new sectors we choose $u=4$, $v=1$
($n=4$, $c=\frac{7}{10}$), which corresponds to the tricritical Ising
model \cite{toro}. One obtains
\begin{equation}\begin{array}{rclclcl}
\D{1}{0} & = & \D{3}{0} & = & \G{1}{0} - \G{0}{1}
& = & (0,0) + (\frac{1}{10},\frac{1}{10})
+ (\frac{3}{5},\frac{3}{5}) + (\frac{3}{2},\frac{3}{2}) \\
\D{2}{0} &   &           & = & \G{2}{0} - \G{0}{2}
& = & 2\cdot (\frac{3}{80},\frac{3}{80}) +
2\cdot (\frac{7}{16},\frac{7}{16})\\
\D{2}{1} & = & \D{2}{3} & = & \G{2}{1} - \G{1}{2}
& = & (\frac{3}{2},0) + (\frac{3}{5},\frac{1}{10})
+ (\frac{1}{10},\frac{3}{5}) + (0,\frac{3}{2})  \\
\D{1/2}{1/2} & = & \D{7/2}{7/2}
& = & \G{1/2}{1/2} - \G{3/2}{3/2}
& = & (0,\frac{7}{16}) + (\frac{1}{10},\frac{3}{80})
+ (\frac{3}{5},\frac{3}{80}) + (\frac{3}{2},\frac{7}{16})\\
\D{7/2}{1/2} & = & \D{1/2}{7/2}
& = & \G{7/2}{1/2} - \G{5/2}{3/2}
& = & (\frac{7}{16},0) + (\frac{3}{80},\frac{1}{10})
+ (\frac{3}{80},\frac{3}{5}) + (\frac{7}{16},\frac{3}{2})
\end{array}\label{3eL20}\end{equation}
which defines a model which differs from the corresponding $R=1$ model
 (cf. eq.~(\ref{3e43})).\vspace{2ex}

\noindent (c)\quad\ul{$v$ even:}\newline At last we again turn to the
case of even value for $v$.  As for the $R=1$ models there are no new
sectors for half-integer values of $\tilde{f}$ in eq.~(\ref{3eL15})
because these do not result in new boundary conditions due to the
identity $(k+\frac{u}{2})\ell_{0} = k\ell_{0}+\frac{v}{2} \equiv
k\ell_{0} \bmod 1$.

\subsubsection{The \mbox{{\boldmath $L=2$}} models}

As above, the discussion of the $L=2$ models is analogous to the $R=2$
models. The unitary subset corresponds to the $O(p)$ models,
$p=2\cos{(\pi/u)}$. Only the integer charge sectors for $u$ odd
($v=1$) were discussed in \cite{toro} ($1_L$ models).  If $u+v$ is
even, eq.~(\ref{3eL11}) applies and from eq.~(\ref{3eL12}) one finds
\begin{eqnarray}
\D{2f}{4g}\zz
\; =\; \D{2(u-f)}{2(u-2g)}\zz
& = & \G{2f}{4g}\zz - \G{2g}{4f}\zz \nonumber \\
& = & \sum_{w=1}^{\frac{u+v}{2}-1}
\C{2(f+g),2w}(z)\,\chi_{2(f-g),2w}(\zquer )  .
\label{3eL21}\end{eqnarray}
Additional sectors are obtained as  follows.\vspace{2ex}

\noindent (a)\quad\ul{$(u+v)$ even:}\newline
One obtains:
\begin{eqnarray}
\G{k}{Q}\zz & = & \;\sum_{d=1-\xi}^{\frac{u+v}{2}-\xi}
\OM{k+\frac{Q}{2}+\eta u}{\xi +2d}(z)\,
\OM{k-\frac{Q}{2}+\eta u}{\xi +2d}(\zquer ) \nonumber \\
& & + \sum_{d=0}^{\frac{u+v}{2}-1}
\OM{k+\frac{Q}{2}+\eta u}{-(\xi +2d)}(z)\,
\OM{k-\frac{Q}{2}+\eta u}{-(\xi +2d)}(\zquer )
\label{3eL22}\end{eqnarray}
where $\eta  =(1-(-1)^{Q})/4$ and $\xi =(1-(-1)^{k})/2$ 
($\xi =(1-(-1)^{k+Q})/2$) for $u+v\equiv 0\bmod 4$ ($u+v\equiv 2\bmod 4$),
respectively.
The following additional sectors appear:
\begin{equation}
\D{2f+\omega}{4g+t}\zz\; =\; \G{2f+\omega}{4g+t}\zz -
\G{\frac{4g+t}{2}+\varepsilon\frac{2\omega +t}{2}u}{2(2f+\omega )+tu}\zz
\label{3eL23}
\end{equation} 
for $\omega\in\{ 0,1\}$, $t\in\{ 0,1,2,3\}$, and $\varepsilon
=(-1)^{(u+1)/2}$.  They are given as a bilinear expression in Virasoro
characters as follows
\begin{equation}
\D{k}{Q}\zz \; = \; \sum_{d=1-\xi}^{\frac{u+v}{2}-1}
\C{k+\frac{Q}{2}+\eta u}{\xi +2d}(z)\,
\C{k-\frac{Q}{2}+\eta u}{\xi +2d}(\zquer )  .
\label{3eL24}\end{equation}
Of course, the character functions that enter in eq.~(\ref{3eL24})
have to comply with the conditions given by eq.~(\ref{3e14}) in order
to obtain physical sectors.\vspace{2ex}

\noindent (b)\quad\ul{$(u+v)$ odd:}\newline Here, one again has to
combine two sectors $\G{k}{Q}\zz$ as in the $R=2$ case
(cf. eq.~(\ref{R2e11})).  The partition functions \mbox{$\GT{k}{Q}\zz
= \G{k}{Q}\zz +\G{k+u}{Q}\zz$} coincide with the sectors
$\G{k}{Q/2}\zz$ (\ref{3eL11}) of the corresponding $L=1$ model with
$n=u$.

This completes our discussion of the projection mechanism in the
finite-size scaling limit.  We now turn our attention to chains of
{\em finite}\/ length $N$ and to their spectra.

\section{Projection mechanism for finite systems}
\setcounter{equation}{0}

We commence this section by explaining in which sense the projection
mechanism which we so far have only established for the continuum
limit, can also be applied to finite systems.  For this purpose we try
to give a meaning to differences of partition functions for a finite
number of sites $N$.

Consider a general sector
\begin{equation}
\D{k}{Q}\zz \; =\;  \G{k}{Q}\zz \; -\;
\G{k^{\prime}}{Q^{\prime}}\zz  .
\label{fe1}\end{equation}
The finite-size analogon of this equation would be
\begin{equation}
\D{k}{Q}(z,\zquer ,N) \;
 = \; \G{k}{Q}(z,\zquer ,N) \; - \;
\G{k^{\prime}}{Q^{\prime}}(z,\zquer ,N)
\label{fe2}\end{equation}
where the $\G{k}{Q}(z,\zquer ,N)$ are defined\footnote{This is true
for minimal models, of course the generic case would be obtained by
replacing $\G{k}{Q}$ by $\E{k}{Q}$ throughout this section.}  by
summing up the finite-size partition functions
\mbox{$\F{k}{Q+n\mu}(z,\zquer ,N)$} (\ref{3e19}) with the appropriate
value of $n$.  We want to interpret the function
$\D{k}{Q}(z,\zquer,N)$ defined by eq.~(\ref{fe2}) as the partition
function of a projected system.  This can be done provided that {\em
any}\/ level which contributes to the partition function
$\G{k^{\prime}}{Q^{\prime}}(z,\zquer ,N)$ has a correspondent in
$\G{k}{Q}(z,\zquer ,N)$, i.e., for any eigenstate of XXZ Heisenberg
chain contributing to $\G{k^{\prime}}{Q^{\prime}}(z,\zquer ,N)$ there
is at least one eigenstate with the same energy and momentum which
contributes to $\G{k}{Q}(z,\zquer ,N)$.  In this case the difference
$\D{k}{Q}(z,\zquer ,N)$ of the two partition functions
$\G{k}{Q}(z,\zquer ,N)$ and $\G{k^{\prime}}{Q^{\prime}}(z,\zquer ,N)$
is the partition function of a system consisting only of those states
which are left over if one eliminates all the degenerate doublets with
one state in each sector.  Denoting in analogy to the notation of
appendix~A (cf. eqs.~(\ref{2e26}) and (\ref{2e36})) the set of all
pairs of energy and momentum eigenvalues\footnote{There is a small
difference to the definitions used in appendix~A since we subtracted a
suitably chosen ground-state energy throughout sec.~2.  For obvious
reasons, however, this does not affect the arguments for degeneracies
of eigenvalues of finite chains.  But note that the index $k$ denotes
different boundary conditions depending on the type of model
considered whereas in appendix~A the boundary condition for
$\GH{K}{Q}(N)$ is given by $\alpha=q^{\, 2K}$ (cf.\
eq.~(\ref{2e20})).}  that contribute to $\G{k}{Q}(z,\zquer ,N)$ by
$\G{k}{Q}(N)$, the condition means that the sets $\G{k}{Q}(N)$ and
$\G{k^{\prime}}{Q^{\prime}}(N)$ should satisfy the inclusion
\begin{equation}
\G{k}{Q}(N) \; \supset \;
\G{k^{\prime}}{Q^{\prime}}(N)  .
\label{fe3}\end{equation}
If this is true, the projection mechanism therefore actually {\em
defines}\/ a finite-size model (in the sense that it determines the
spectrum) as long as we are in the physical region $h \geq 1/4$
(\ref{3e3}) of the XXZ Heisenberg chain. This condition is fulfilled
for all $R$-models and for the $L>1$ models where \mbox{$h \geq 1/4$}
if \mbox{$m \geq 1/(L^{2}-1)$}, but it for instance excludes all $L=1$
models.

In appendix~A, we establish intertwining relations between charge
sectors of the XXZ Hamiltonian with different toroidal boundary
conditions using powers of the quantum algebra generators of \UQSU\
(the corresponding representation on ${\cal H}(N)$ is given by
eq.~(\ref{2e14})).  Analogous intertwining relations hold for the
corresponding translation operators (\ref{2e5}).  These relations
allow us to obtain inclusion relations for the sets of simultaneous
eigenvalues of the Hamiltonian and the translation operator of the
form (\ref{fe3}). The main results obtained in appendix~A are
eq.~(\ref{2e30}) for generic values of $q$ and eq.~(\ref{2e35}) if $q$
is a root of unity.  We proceed by having a closer look at the $R$-
and $L$-models separately.

\subsection{The \mbox{\protect\boldmath $R$}-models}

For the $R$ models, the connection between the anisotropy $\gamma$ and
the boundary condition is determined by the equations
(cf. eqs.~(\ref{3e20}) and (\ref{3e22}))
\begin{eqnarray}
q & = & -\exp\left(-i\pi\gamma\right)
 \; = \; \exp\left( i \pi\frac{m}{R^{2} (m+1)}\right) \nonumber \\
\alpha (k)
& = & \exp \left( 2\pi i k (\ell_{0} + \nu_{0}) \right)
\; = \; \exp \left( \frac{2\pi i k}{R (m+1)} \right)
\label{fe5}\end{eqnarray}
where $m$ and $k$ are arbitrary. It follows that
\begin{equation}
\alpha (k) \; = \; e^{\, 2 \pi i k/R} \; q^{\, -2 R k}  .
\label{fe6}\end{equation}
The phase factor in eq.~(\ref{fe6}) is equal to one if $k$ is an
integer multiple of $R$.  In this case one obtains with $k = R f$, $f$
integer, the following relation
\begin{equation}
\alpha (R f) \; = \; q^{\, -2 R^2 f}  ,
\label{fe7}\end{equation}
i.e., if $k = R f$ then the corresponding value of $K$ in the notation
of appendix~A would be \mbox{$K = -R^{2} f = - R k$}.  Comparing now
the inclusion relations given in eqs.~(\ref{2e30}) and (\ref{2e35})
(choosing \mbox{$K = -R^{2} f$} and \mbox{$Q = R^2 g$} with integers
$f$ and $g$ fulfilling \mbox{$0 \leq |g| \leq |f|$}) with the sectors
(\ref{3e24}) respective (\ref{3e36}), one immediately realizes that
for all $R$-models and all the sectors considered the inclusion
relations (\ref{fe3}) is fulfilled and thus the projection mechanism
extends to finite chains.

We now focus on the minimal case, i.e., $m=u/v$ with positive coprime
integers $u$ and $v$.  In sec.~2, we explicitly obtained additional
sectors for the $R=1$ (see sec.~2.1.1) and the $R=2$ (see sec.~2.1.2)
models.  In ref.~\cite{toro} it was observed numerically that in the
case of the unitary minimal series (i.e., $v=1$) for the $R=2$
models\footnote{Remember that these are the $1_{R}$ models in the
notation of ref.~\cite{toro}.}  only part of the sectors given in
eqs.~(\ref{R2e5})--(\ref{R2e6}) show the degeneracies (\ref{fe3}). In
particular, this is true for all sectors $\D{k}{Q}\zz$ of
eq.~(\ref{R2e6}) with \mbox{$\xi = 0$}, which are the sectors
$\D{k}{Q}\zz$ with even $k$ (even $k+Q$) for \mbox{$u \equiv 0 \bmod
4$} (\mbox{$u \equiv 2 \bmod 4$}), respectively.

To understand these observations, let us consider values of $u$ which
are multiples of $R$, i.e., \mbox{$u \equiv 0 \bmod R$}. This gives
$q^{\pm R(u+v)} = (-1)^{u/R}$ and hence Eq.~(\ref{fe6}) can be
modified as follows
\begin{equation}
\alpha (k) \; = \; (-1)^{\, u/R}\; e^{\, 2 \pi i k/R}
\; q^{\, -2\, (R k \,\pm\, R\, (u+v)/2)}  .
\label{fe9}\end{equation}
This means that if \ul{$u/R$ is even}, i.e., \mbox{$u \equiv 0 \bmod
2R$}, one obtains as a generalization of eq.~(\ref{fe7}) for \mbox{$k
= R \tilde{f}$}, $\tilde{f}$ integer, the following equation
\begin{equation}
\alpha (R \tilde{f}) \; = \; q^{\, -2 R^2 \tilde{f}} \; = \;
q^{\, -2\, (R^2 \tilde{f} \,\pm\, R\, (u+v)/2)}  .
\label{fe10}\end{equation}
Analogously, if \ul{$u/R$ is odd}, i.e., \mbox{$u \equiv R \bmod 2R$},
one obtains with \mbox{$k = R \tilde{f}$} in addition to
eq.~(\ref{fe7}), but now for {\em half-integer}\/ values of
$\tilde{f}$ (i.e., $2\tilde{f}$ is an {\em odd}\/ integer), the
relation
\begin{equation}
\alpha (R \tilde{f}) \; = \; -\, q^{\, -2 R^2 \tilde{f}} \; = \;
q^{\, -2\, (R^2 \tilde{f} \,\pm\, R\, (u+v)/2)}  .
\label{fe11}\end{equation}
The inclusion relation (\ref{2e35}) with \mbox{$K = -R^2 \tilde{f} \pm R (u+v)/2$}
and \mbox{$Q = R^2 \tilde{g} \pm R (u+v)/2$} now result in the
following relations
\begin{equation}
\G{R \tilde{f}}{R^2 \tilde{g} \,\pm\, R\, (u+v)/2}(N) \; \supset \;
\G{R \tilde{g}}{R^2 \tilde{f} \,\pm\, R\, (u+v)/2}(N)
\label{fe12}\end{equation}
with appropriate values of $\tilde{f}$ and $\tilde{g}$.  From this one
deduces that in fact {\em all}\/ the additional sectors (see
eqs.~(\ref{3e41}) and (\ref{3e44})) for the $R=1$ models show the
degeneracies (\ref{fe3}), whereas for the $R=2$ models this is true
for all sectors $\D{k}{Q}\zz$ (\ref{R2e6}) with even values of $k$ if
\mbox{$u \equiv 0 \bmod 4$} respective for all sectors $\D{k}{Q}\zz$
(\ref{R2e6}) with even $k+Q$ if \mbox{$u \equiv 2 \bmod 4$} (for
arbitrary value of $v$).

\subsection{The \mbox{\protect\boldmath $L$}-models}

The discussion of the $L$-models is completely analogous to the
previous section on the $R$-models.  One should keep in mind that
results on the finite-chain spectra only apply if $h \geq 1/4$
(\ref{3e3}).  The connection between the anisotropy $\gamma$ and the
boundary condition is given by (\ref{3eL3})
\begin{eqnarray}
q & = & -\exp\left(-i\pi\gamma\right)
 \; = \;\exp\left( i \pi\frac{m+1}{L^{2} m}\right) \nonumber \\
\alpha (k) 
& = & \exp\left( 2\pi i k (\ell_{0} + \nu_{0}) \right)
\; = \;\exp \left( \frac{2\pi i k}{L m} \right)
\label{feL2}\end{eqnarray}
and hence
\begin{equation}
\alpha (k) \; = \; e^{\, -2 \pi i k/L} \; q^{\, 2 L k}  .
\label{feL3}\end{equation}
Considering values of $k$ which are integer multiples of $L$,
i.e., $k = L f$, $f$ integer, one obtains
\begin{equation}
\alpha (L f) \; = \; q^{\, 2 L^2 f}  ,
\label{feL4}\end{equation}
i.e., the corresponding value of $K$ in the notation of appendix~A
would be \mbox{$K = L^{2} f = L k$}.  Again, the inclusion relations
(\ref{2e30}) and (\ref{2e35}), now with the choices \mbox{$K = L^{2}
f$} and \mbox{$Q = L^2 g$} ($f$, $g$ integers fulfilling \mbox{$0 \leq
|g| \leq |f|$}), guarantee that all for all sectors given by
eqs.~(\ref{3eL5}) respective (\ref{3eL12}) the projection mechanism
can be applied for finite-size systems.

Finally, we again consider the minimal case $m=u/v$, where $u$ and $v$
are coprime positive integers. Here, we consider models where
\mbox{$u+v \equiv 0 \bmod L$}. Using $q^{\pm Lu} =(-1)^{(u+v)/L}$, one
can modify eq.~(\ref{feL3}) as follows
\begin{equation}
\alpha (k) \; = \; (-1)^{\, (u+v)/L} e^{\, -2 \pi i k/L}
\; q^{\, 2\, (L k \pm L u/2)}  .
\label{feL6}\end{equation}
If \ul{$(u+v)/L$ is even}, i.e., \mbox{$u+v \equiv 0 \bmod 2L$}, one
obtains for \mbox{$k = L \tilde{f}$}, $\tilde{f}$ integer, the
expression
\begin{equation}
\alpha (L \tilde{f}) \; = \; q^{\, 2 L^2 \tilde{f}} \; = \;
q^{\, 2\, (L^2 \tilde{f} \pm L u/2)}  ,
\label{feL7}\end{equation}
whereas if \ul{$(u+v)/L$ is odd}, i.e., \mbox{$u+v \equiv L \bmod 2L$},
substituting \mbox{$k = L \tilde{f}$}
with {\em half-integer}\/ values of $\tilde{f}$ into eq.~(\ref{feL6}) yields
\begin{equation}
\alpha (L \tilde{f}) \; = \; -\, q^{\, 2 L^2 \tilde{f}} \; = \;
q^{\, 2\, (L^2 \tilde{f} \pm L u/2)}  .
\label{feL8}\end{equation}
We now use the inclusion relation (\ref{2e35}) with
\mbox{$K = L^2 \tilde{f} \pm L u/2$} and
\mbox{$Q = L^2 \tilde{g} \pm L u/2$} and obtain the relation
\begin{equation}
\G{L \tilde{f}}{L^2 \tilde{g} \,\pm\, L u/2}(N) \; \supset \;
\G{L \tilde{g}}{L^2 \tilde{f} \,\pm\, L u/2}(N)
\label{feL9}\end{equation}
where again $\tilde{f}$ and $\tilde{g}$ have to be chosen
appropriately.  For the $L=2$ models (note that the $L=1$ models all
have $h<1/4$), one finds the following behaviour.  The sectors
$\D{k}{Q}\zz$ in Eq.~(\ref{3eL24}) possess the finite-size
degeneracies (\ref{fe3}) if $\xi = 0$, that means if $k$ ($k+Q$) is
even for \mbox{$u+v \equiv 0 \bmod 4$} (\mbox{$u+v \equiv 2 \bmod
4$}), respectively.  This proves and extends the numerical results of
ref.~\cite{toro}.

Let us summarize the results of this section.  We explicitly derived
all the degeneracies observed numerically in ref.~\cite{toro} using
the results of appendix~A. In fact, there is a lot more information in
our equations. In particular, we could show that for the $R=1$ models
all sectors (including those with half-integer charges) show the
required degeneracies (\ref{fe3}) to define a finite-size model.  In
ref.~\cite{toro}, it has been shown (by numerical comparison) that the
spectrum of the $R=1$ model with $u=3$, $v=1$ (central charge $c=1/2$)
is exactly that of the Ising quantum chain and that the $R=1$ model
with $u=5$, $v=1$ ($c=4/5$) reproduces the spectrum of the 3-states
Potts quantum chain (see \cite{GR86} and references therein), in both
cases with toroidal boundary conditions.  In the subsequent section,
we are going to address the question what the new sectors in these
$R=1$ models correspond to in the Ising respective 3-states Potts
quantum chain.

\section{Interpretation of the new sectors}
\setcounter{equation}{0}

\subsection{New boundary conditions for the Ising and 
3-states Potts quantum chains}

In this section we discuss the significance of the new half-integer
charge sectors in the projected systems and show that they are related
to a new kind of boundary condition in these models.  Before we do so
we want to make the problem at hand more precise by reminding the
reader of the relation between the labels $k$ and $q$ in the partition
functions $\D{k}{q}$ and the projected systems they correspond to.

As an example consider the two-dimensional Ising model on a torus.  In
the extreme anisotropic limit it is described by the lattice
Hamiltonian \cite{Kogut}
\begin{equation}
\label{Ising}
H\; =\; -\,\frac{1}{2}\;
\sum_{j=1}^{M} \;\left(\: \sx{j}\; +\; \lambda\: \sz{j} \sz{j+1} \:\right)
\end{equation}
Here, $M$ represents the number of sites and $\lambda$ plays the role
of the inverse temperature.  In the thermodynamic limit $M \rightarrow
\infty$ the model has a critical point at $\lambda=1$ where it is
described by a conformal field theory with central charge $c=1/2$ (a
Majorana fermion).  The Ising model has a global $\ZZ_{\, 2}$-symmetry
(the spin-flip operation), as a consequence $H$ commutes with the
operator
\begin{equation}
S\; =\; \prod_{j=1}^{M}\: \sx{j}
\label{Flip}
\end{equation}
with eigenvalues $S=\pm 1$ splitting $H$ into two sectors, one of
which is even under this operation ($S=1$), the other one odd
($S=-1$).  On the other hand it is known that for quantum chains whose
group of global symmetries is of order $n$, there are also $n$
different types of toroidal boundary conditions, i.e., boundary
conditions compatible with the geometry of a torus \cite{Chas}. Here
$n=2$ and one has periodic boundary conditions ($\sz{M+1} = \sz{1}$)
and antiperiodic boundary conditions ($\sz{M+1} = - \sz{1}$).  In the
two-dimensional model from which $H$ is obtained the latter correspond
to a seam of antiferromagnetic bonds in an otherwise ferromagnetic
system. The choice of boundary conditions is reflected in the
structure of the translation operator in a very intuitive way: With
periodic boundary conditions, (\ref{Ising}) commutes with the
translation operator
\begin{equation}\label{TISP}
T\; =\; \OP{j=1}{M-1} P_{j}  .
\end{equation}
with $P_j$ defined in (\ref{2e5}).  On the other hand, in the case of
antiperiodic boundary conditions an additional spin-flip operation at
the boundary is necessary in order to construct a commuting
translation operator $T'$:
\begin{equation}\label{TISA}
T'\; =\;  T\,\sx{M}  .
\end{equation}
One finds $T^{M}=1$ and ${T'}^{M}=S$.

To conclude this short reminder of the Ising quantum chain let us
denote the eigenvalues of (\ref{Ising}) by $\EPS{l'}{l}{j}$, $l=0$
($l=1$) in the even (odd) sector, $l'=0$ ($l'=1$) for periodic
(antiperiodic) boundary conditions and $j=1,\ldots ,2^{M-1}$ according
to the number of states in these sectors.

As already noted, the numbers $k$ and $q$ in the partition functions
$\D{k}{q}$ of the projected systems label the sectors of these models
according to their internal global symmetries and the type of toroidal
boundary condition imposed on the systems.  The discussion of the
$R=1$ model with $u=3$ and $v=1$ in sec.~2 in the thermodynamic limit
$M\rightarrow\infty$ amounts to the statement that at the critical
point \mbox{$\lambda = 1$} the projected sectors (\ref{3e45a}) of the
XXZ Heisenberg chain coincide with the sectors described here (in the
limit $N\rightarrow\infty$).  In particular, the levels contributing
to $\D{1}{0}$ are the scaled energy gaps $\EPSquer{0}{0}{j}$ (here
$\EPSquer{l'}{l}{j} = M/2\pi (\EPS{l'}{l}{j} - \EPS{0}{0}{1})$), to
$\D{2}{0}$ contribute $\EPSquer{0}{1}{j}$ and $\EPSquer{1}{0}{j}$
which are degenerate, and to $\D{2}{1}$ contribute
$\EPSquer{1}{1}{j}$.

This raises the question: To which boundary conditions and sectors of
the Ising model correspond the sectors $\D{1/2}{1/2}$ and
$\D{7/2}{1/2}$ of eq.~(\ref{3e45a})?  The existence of these sectors
is actually surprising as the spin-flip symmetry allows only for the
periodic and antiperiodic toroidal boundary conditions for $H$
discussed above and therefore no new toroidal boundary conditions
should be expected. This observation allows us to formulate the
problem: One has to find an additional symmetry of the Ising model and
the corresponding boundary conditions which give the spectrum
corresponding the sectors $\D{1/2}{1/2}$ and $\D{7/2}{1/2}$ of
(\ref{3e45a})\footnote{The Ising Hamiltonian (\protect\ref{Ising}) is
also invariant under a parity operation, but this is of no importance
to the present discussion.}.

In order to answer this question we note that even for {\em finite}\/
systems the scaled energy gaps $\EPSquer{l'}{l}{j}$ of the Ising
Hamiltonian with $M$ sites are identical to those scaled energy gaps
of the XXZ Heisenberg chain with $N=2M$ sites which contribute to the
integer charge sectors $\D{k}{q}$ of (\ref{3e45a}).  The reason for
this important observation is studied in refs.~\cite{TLA,DL,GS} and we
do not want to repeat the discussion here. In ref.~\cite{GS}, we
construct lattice Hamiltonians for the Ising and $3$-state Potts
models which have the spectrum contributing to the new partition
functions of the XXZ Heisenberg chain with $N=2M-1$ sites (see below).
Given these Hamiltonians, we can analyze their symmetries and the
physical significance of the boundary conditions involved.

\subsubsection{Ising model}

A Hamiltonian such that its scaled energy gaps coincide with those
obtained through the projection mechanism from the normalized XXZ
Heisenberg chain (\ref{3e2}) with $q=-\exp{(i \pi/4)}$, $\alpha = -
q^{-2k}$ and $N=2M-1$ sites\footnote{The scaling factor $N/2\pi$ in
(\protect\ref{3e7}) has to be changed into $M/\pi$ if $N=2M-1$.} is
given by \cite{GS}
\begin{equation}\label{HIS}
\tilde{H}\; =\; -\sum_{j=1}^{2M-1} \left(e_{\, j} - \frac{1}{2}\right)
\end{equation}
with
\begin{eqnarray}
\label{TLAIS2}
e_{\, 2j-1} & = & \frac{1}{2}\, (1 + \sx{j})  ,\qquad 
e_{\, 2j}   \; = \; \frac{1}{2}\, (1 + \sz{j}\sz{j+1}) \\
e_{\, 2M-1}^{\,\pm} & = & \displaystyle\frac{1}{2}\, (1 \pm \sy{M}\sz{1})  .
\label{TLAIS2b}\end{eqnarray}
where $1\leq j\leq M\! -\! 1$.  $\tilde{H}$ is an Ising Hamiltonian
acting on a chain of $M$ sites with a boundary term $\pm
\sy{M}\sz{1}$.  Note that as opposed to periodic and antiperiodic
boundary conditions as well as to the ``generalized defects''
investigated in ref.~\cite{uwe} (which include boundary couplings of
the form $\sy{M}\sz{1}$) there is no operator $\sx{M}$ present. To the
best of our knowledge this type of boundary condition has not yet been
studied in the literature. The spectrum of $\tilde{H}$ leads to the
partition functions (\ref{3e45a}) for half-integer charge sectors. The
highest weight representations of the Virasoro algebra (0,1/16),
(1/16,0), (1/2,1/16) and (1/16,1/2) contributing to these partition
functions represent the anomalous dimensions of some new spinor
fields.

The two different signs in the boundary term amount to complex
conjugation. Since the Hamiltonian (\ref{HIS}) is hermitian, the
corresponding spectra are identical. This is indeed observed in the
projected spectra: The eigenvalues of the XXZ Heisenberg chain
contributing to the sectors $\D{1/2}{1/2}$ and $\D{7/2}{1/2}$ resp.
are related through complex conjugation of the boundary angle $\alpha$
in (2.1). Since for the choice $q$ and $\alpha$ under consideration
the XXZ Heisenberg chain is hermitian and all eigenvalues therefore
real, the two spectra are degenerate.

In order to understand the symmetry from which the new boundary
conditions arise we go back to the standard Ising model defined by the
Hamiltonian (\ref{Ising}).  First we want to stress again that the
eigenvalues of $H$ (\ref{Ising}) with $M$ sites and $\lambda =1$ are
obtained through the projection mechanism from the XXZ Heisenberg
chain (2.1) with $N=2M$ sites. The eigenstates of the XXZ Heisenberg
chain are also eigenstates of the translation operator (\ref{2e5})
which has $2M$ different eigenvalues of the form $\exp{((2\pi i k + i
\pi\phi)/2M)}$, $0 \leq k \leq 2M-1$. As shown in appendix A the
projected eigenstates of (2.1) remain eigenstates also of
(\ref{2e5}). On the other hand the translation operators (\ref{TISP})
and (\ref{TISA}) of the Ising model have only $M$ different
eigenvalues of the form $\exp{((2\pi i k + i \pi l)/M)}$, $0\leq k
\leq M-1$, (with $l=0,1$ depending on the sector). From this
observation we learn that $T$ and $T'$ of the Ising model
(\ref{Ising}) are {\em not}\/ the equivalent operators to the
translation operator $T(\alpha,2M)$ of the XXZ Heisenberg chain but
that they correspond to $T^{\, 2}(\alpha,2M)$. Technically speaking
this means that on the projected subspace of the XXZ Heisenberg chain
under consideration the translation operators $T$ and $T'$ are not
representations of the translation operator $T(\alpha,2M)$ but of its
square $T^{\, 2}(\alpha,2M)$\footnote{This explains why we had to
neglect macroscopic momenta $\pi$ in the spectrum of the XXZ
Heisenberg chain to get the momenta in the Ising chain}. But since the
projected eigenstates are also eigenstates of $T(\alpha,2M)$ we have
established the presence of an additional symmetry in the Ising model
besides the $\ZZ_{\, 2}$ spin-flip symmetry and translational
invariance.

The physical meaning of the symmetry generated by $T(\alpha,2M)$ in
the projected system can be understood by studying its representation
$D$ in the Ising model. One finds that $D$ satisfies \cite{DL,GS}
\begin{equation}
D\; e_{\, j'} \; =\; e_{\, j'+1}\; D
\end{equation}
where the $e_{\, j'}$, $1 \leq j' \leq 2M$, are defined as in
eq.~(\ref{TLAIS2}) by extending the range of $j$ to $1\leq j\leq M$.
This is the duality transformation $D$ (see appendix~B).  We conclude
that the additional symmetry we found is duality which at (and only
at) the self-dual point $\lambda =1$ is indeed a true symmetry: If
$\lambda=1$ the duality relation (\ref{dual}) becomes
\begin{equation}\begin{array}{ccccc}
H^D(1) & = & D H(1) D^{-1} & = & H(1)
\end{array}\end{equation}
Since $D^{\, 2}$ corresponds to translations (see appendix~B) we can
say that duality is the ``square root'' of translations. Obviously,
not only the standard Ising Hamiltonian (\ref{Ising}) but also
$\tilde{H}$ (\ref{HIS}) commute with the duality transformation (in an
appropriate representation $\tilde{D}$)\footnote{Strictly speaking,
this statement applies to the mixed sector versions of
(\protect\ref{Ising}) and (\protect\ref{HIS}) discussed in
appendix~B. Here we have omitted all subscripts and superscripts
specifying the various sectors and boundary conditions in the Ising
model and hence the representation of $D$. A detailed discussion is
given in appendix~B.}.

Having found an additional symmetry in the Ising model at its
self-dual point and the Hamiltonian (\ref{HIS}) giving rise to the new
sectors discovered through the projection mechanism we can proceed
looking for an explicit representation of the translation operator
commuting with (\ref{HIS}). As in the standard case of periodic and
antiperiodic boundary conditions this sheds light on the physical
meaning of the boundary conditions.  One finds corresponding to the
two possible choices of the sign in $e_{\, 2M-1}$ the translation
operator $\tilde{T}$ and its complex conjugate $\tilde{T}^{\ast}$
commuting with $\tilde{H}$ and $\tilde{H}^{\ast}$ resp.
\begin{equation} \label{hatt}
\tilde{T}\; =\; T\: g_{\, 2M-2}\: g_{\, 2M-1}
\end{equation}
with $T$ given in (\ref{TISP}). The operators $g_{\, j}$ are related to
the duality transformation (\ref{dut}) and defined by
\begin{equation}
\label{gg}
 g_{\, 2M-2} \; = \;  - \,\frac{1-i}{2}\, \left( 1 - i \sz{M-1} \sz{M}
\right),\qquad
 g_{\, 2M-1} \; = \; - \,\frac{1-i}{2}\, \left( 1 - i \sx{M} \right).
\end{equation}
A straightforward calculation shows that the $M^{\mbox{th}}$ power of
$\tilde{T}$ is the duality transformation in each sector (see
appendix~B).  This clarifies the meaning of this type of boundary
condition: $\tilde{T}$ performs the local equivalent of the duality
transformation at the boundary in addition to a pure translation.  The
$M^{\mbox{th}}$ power of $\tilde{T}$ gives the symmetry operator to
which the boundary condition is related.  This exhibits its relation
to the duality symmetry in the same way as the existence of the
spin-flip symmetry resulted in the existence of the translation
operator $T'$ acting locally at the boundary as spin-flip operator
times a global translation (cf. also the translation operator
(\ref{2e5}) for the XXZ chain with toroidal boundary conditions).

\subsubsection{3-state Potts model}

The $3$-state Potts model is obtained from the $R=1$ series with $u=5$
and $v=1$. The discussion of the new boundary conditions here is in
complete analogy to the previous discussion of the Ising model and we
state only the results.

Taking the extreme anisotropic limit of the transfer matrix of the
$3$-state Potts model on a torus with periodic boundary conditions one
obtains the Hamiltonian \cite{GR86}
\begin{equation}\label{HP3}
H\; =\; - \frac{2}{3\sqrt{3}}\; \sum_{j=1}^{M}\; \Gamma_{\, j} +
\Gamma_{\,j}^{\, 2} + \lambda\, ( \sigma_{\, j}\sigma_{\, j+1}^{\, 2}
+ \sigma_{\, j}^{\,2}\sigma_{\, j+1} )
\end{equation}
Here, $\Gamma_{\, j}$ and $\sigma_{\, j}$ are the matrices
\renewcommand{\arraystretch}{0.75}
\begin{equation}
\Gamma_{\, j}\; =\; \left( 
\begin{array}{ccc}  0 &  0 & 1 \\
          1 &  0 & 0 \\
          0 &  1 & 0 \end{array} \right)_{\, j} ,\qquad
\sigma_{\, j}\; =\; \left( 
\begin{array}{ccc} 1 & 0 & 0 \\
         0 & \omega & 0 \\
         0 & 0 & \omega^{2} \end{array} \right)_{\, j}
\end{equation}
\renewcommand{\arraystretch}{1} acting on site $j$ and $\omega =
\exp{(2\pi i/3)}$.  Again $\lambda$ plays the role of the inverse
temperature and in the thermodynamic limit $M \rightarrow \infty$ the
model has a critical point at $\lambda =1$ with central charge
$c=4/5$. The Hamiltonian (\ref{HP3}) with periodic boundary conditions
$\sigma_{\, M+1}=\sigma_{\, 1}$ is symmetric under the permutation
group $S_3$ and commutes with the operators $Z$ and $E$ defined by
\renewcommand{\arraystretch}{0.75}
\begin{equation}
Z\; =\;\prod_{j=1}^{M} \Gamma_{\, j} ,  
E\; =\;\prod_{j=1}^{M} V_{\, j}  ,
V_{\, j}\; =\; \left( 
\begin{array}{ccc} 1 & 0 & 0 \\
         0 & 0 & 1 \\
         0 & 1 & 0 \end{array} \right)_{\, j}  .
\end{equation}
\renewcommand{\arraystretch}{1} They satisfy $Z^{\, 3}=E^{\, 2}=1$ and
$EZ= Z^{\, 2}E$. $H$ splits into four sectors $H_{0,+}$, $H_{0,-}$,
$H_{1}$ and $H_{2}$ corresponding to the four irreducible
representations of $S_3$. They are labelled according to the
eigenvalues $\omega^{\kappa}$, $\kappa=0,1,2$, of $Z$ and $\pm 1$ of
$E$, respectively.

According to the $S_3$-symmetry there are non-periodic toroidal
boundary conditions \cite{Chas} accounting for the various integer
charge sectors (\ref{3e45b}) obtained from the XXZ Heisenberg chain
with $N=2M$ sites.  At the self-dual point $\lambda =1$ the symmetry
is enhanced as the duality transformation becomes a true symmetry of
the (mixed sector) model (see appendix~B).

A Hamiltonian for the $3$-state Potts model such that its scaled energy gaps
coincide with those obtained through the projection mechanism from the
normalized XXZ Heisenberg chain (\ref{3e2}) with $q=-\exp{(i \pi/6)}$,
$\alpha = - q^{-2k}$ and $N=2M-1$ sites is given by \cite{GS}
\begin{equation}\label{3P2}
\tilde{H}\; =\;
-\frac{2}{\sqrt{3}}\sum_{j=1}^{2M-1}\left(e_{\, j}-\frac{1}{3}\right)
 .
\end{equation}
Here, the operators $e_{\, j}$ are defined by
\begin{eqnarray}
e_{\, 2j-1} & = & \displaystyle \frac{1}{3}\, (1 + \Gamma_{\, j} +
\Gamma_{\, j}^{\, 2})   ,  
e_{\, 2j} \; = \;\frac{1}{3}\, 
(1 + \sigma_{\, j}\sigma_{\, j+1}^{\, 2} + 
\sigma_{\, j}^{\, 2}\sigma_{\, j+1}) 
\nonumber \\
e_{\, 2M-1}^{\, (\kappa)} & = & \frac{1}{3}\, (1 +
\omega^{-\kappa}\Gamma_{\,M}^{\,\kappa}\sigma_{\, M}\sigma_{\, 1}^{\, 2} +
\omega^{\kappa}\sigma_{\, M}^{\, 2}\Gamma_{\, M}^{\, -\kappa}\sigma_{\, 1})
\label{P3TLA2}
\end{eqnarray}
where $1\leq j\leq M\! -\! 1$ and $\kappa =1,2$.  Note that $e_{\,
2M-1}^{\, (1)}$ is the complex conjugate of $e_{\, 2M-1}^{\, (2)}$.
Since the Hamiltonian $\tilde{H}$ (\ref{3P2}) is hermitian, the
spectra for $\kappa=1$ and $\kappa=2$ are identical.  This corresponds
to the degeneracy of the energy levels contributing to the partition
functions $\D{1/2}{1/2}$ and $\D{11/2}{1/2}$, $\D{3/2}{1/2}$ and
$\D{9/2}{1/2}$ and $\D{1/2}{3/2}$ and $\D{11/2}{3/2}$ (see
eq.~(\ref{3e45b})).  These pairs of partition functions are obtained
through complex conjugation of the boundary angle $\alpha$ of the XXZ
Heisenberg chain which is also hermitian. Furthermore, analysis of the
number of eigenstates in each sector of $\tilde{H}$ for finite values
of $M$ \cite{spin1,GS} shows that $\tilde{H}_{0,+}$ contains the
energy levels contributing to $\D{3/2}{1/2}$, that $\tilde{H}_{0,-}$
contains the energy levels contributing to $\D{1/2}{3/2}$ and that
$\tilde{H}_{1}\cong \tilde{H}_{2}$ contains the energy levels
contributing to $\D{1/2}{1/2}$.  In the $3$-state Potts model the new
boundary conditions are related to the duality transformation in the
same way as in the Ising model.

\subsection{Numerical Results for the \mbox{\protect\boldmath $R=\,$}1
model with \mbox{\protect\boldmath $u=\,$}2 and
\mbox{\protect\boldmath $v=\,$}3}

As already mentioned in sec.~2, the minimal $R=1$ models with $u=2$
have a special feature: one obtains the same character functions by
the projection procedure described in sec.~2 applied to the XXZ chain
with an even or with an odd number of sites (see eqs.~(\ref{3e37}) and
(\ref{3e41})).  This behaviour is obviously different from the one
observed in the cases of the Ising model ($u=3$, $v=1$) and the
$3$-state Potts model ($u=5$, $v=1$) above.  It is our conjecture that
for these models it is not necessary to distinguish between spectra
obtained from an even or an odd number of sites. The same phenomenon
has been observed for the projection mechanism for open boundary
conditions \cite{free3}.  At this place we want to present some
numerical results for the simplest example of this series, namely the
model with $v=3$ ($n=5$) with central charge $c=-22/5$. The sectors
for this model are given in eq.~(\ref{3e42}).

First we discuss the ground-state energies.  As ground state of our
projected model we choose \mbox{$E_{\,
0\,}(N)=\EE{1}{0}{2}(\frac{3}{5},N)$} for an even number of sites $N$
and \mbox{$E_{\, 0\, }(N)=\EE{0}{3/2}{1}(\frac{3}{5},N)$} for an odd
number of sites $N$. These are the states which in the finite-size
scaling limit give a contribution of $1$ to the partition functions
$\G{1}{0}\zz$ and $\G{0}{3/2}\zz$ for even and odd number of sites,
respectively (see eq.~(\ref{3e42})). In Table~\ref{tab1}, we list the
numerical values of \mbox{$-E_{\, 0\, }(N)/N$} for up to 18
sites. Also shown are the differences between these values and the
first two terms in an $1/N$ expansion. The constant term $A_{0}\approx
0.365315$ is the infinite-size limit of \mbox{$-E_{\, 0\, }(N)/N$}
which was computed from the exact solution \cite{Yangs} by numerical
integration. The form of the second term follows from conformal
invariance \cite{BCN,Aff} and involves the central charge $c=-22/5$ of
our projected model.  Apparently, there is no visible difference in
the behaviour of the ground-state energies for even or odd lengths of
the chain.

{\small
\begin{table}[tb]
\caption{\label{tab1} \protect\footnotesize Ground-state energy per
site of the $R=1$ model with $u=2$, $v=3$ and chains with up to 18
sites}
\begin{center}
\begin{tabular}{|c|r@{.}l|c|}
\hline
$N$ \rud & \multicolumn{2}{|c}{ $-\frac{E_{0}(N)}{N}$} &
\multicolumn{1}{|c|}{$-\frac{E_{0}(N)}{N}-A_{0}-\frac{\pi c}{6N^{2}}$} \\
\hline
\ru $2$ & $-0$&$292\, 428$ & $-0.081\, 784$ \\
    $3$  & $0$&$097\, 476$ & $-0.011\, 858$ \\
    $4$  & $0$&$217\, 963$ & $-0.003\, 362$ \\
    $5$  & $0$&$271\, 846$ & $-0.001\, 316$ \\
    $6$  & $0$&$300\, 700$ & $-0.000\, 620$ \\
    $7$  & $0$&$317\, 968$ & $-0.000\, 330$ \\
    $8$  & $0$&$329\, 126$ & $-0.000\, 192$ \\
    $9$  & $0$&$336\, 753$ & $-0.000\, 119$ \\
    $10$ & $0$&$342\, 199$ & $-0.000\, 078$ \\
    $11$ & $0$&$346\, 222$ & $-0.000\, 053$ \\
    $12$ & $0$&$349\, 279$ & $-0.000\, 037$ \\
    $13$ & $0$&$351\, 656$ & $-0.000\, 027$ \\
    $14$ & $0$&$353\, 541$ & $-0.000\, 020$ \\
    $15$ & $0$&$355\, 061$ & $-0.000\, 015$ \\
    $16$ & $0$&$356\, 304$ & $-0.000\, 012$ \\
    $17$ & $0$&$357\, 334$ & $-0.000\, 009$ \\
\rd $18$ & $0$&$358\, 197$ & $-0.000\, 007$ \\
\hline
\end{tabular}
\end{center}
\end{table}
}

{\small
\begin{table}[tb]
\caption{\label{tab2} \protect\footnotesize
The lowest scaled energy gaps that contribute to the
sectors $\D{1}{0}\zz$ resp. $\D{0}{3/2}\zz$
and $\D{2}{1}$ resp. $\D{4}{1/2}$ are given
for chains of length $N$. Also shown are
corresponding exact values for $N \rightarrow \infty$
and the differences $\delta$ between the exact and the
extrapolated values (see text for details).}
\begin{center}
\begin{tabular}{|r||l|l||l|l|l|l|}
\hline
\multicolumn{1}{|c||}{\rud $N$} &
\multicolumn{2}{|c||}{$\C{1}{1}(z)\,\C{1}{1}(\zquer )$} &
\multicolumn{4}{|c|}{$\C{1}{1}(z)\,\C{1}{2}(\zquer )$} \\
\hline
\ru
 2 &                  &                  & -0.248\, 220 &             &
               &             \\
 3 &                  &                  & -0.216\, 948 & 0.775\, 444 &
               &             \\
 4 & \ul{1.606\, 516} &                  & -0.208\, 906 & 0.803\, 258 &
   1.318\, 981 &             \\
 5 & \ul{1.816\, 384} &                  & -0.205\, 547 & 0.808\, 298 &
   1.554\, 938 & 1.565\, 539 \\
 6 & \ul{1.909\, 047} & \ul{2.281\, 378} & -0.203\, 801 & 0.808\, 328 &
   1.664\, 745 & 1.678\, 934 \\
 7 & \ul{1.952\, 922} & \ul{2.552\, 745} & -0.202\, 771 & 0.807\, 313 &
   1.719\, 916 & 1.734\, 239 \\
 8 & \ul{1.975\, 088} & \ul{2.712\, 909} & -0.202\, 111 & 0.806\, 210 &
   1.749\, 707 & 1.762\, 924 \\
 9 & \ul{1.986\, 887} & \ul{2.810\, 122} & -0.201\, 662 & 0.805\, 245 &
   1.766\, 826 & 1.778\, 613 \\
10 & \ul{1.993\, 425} & \ul{2.870\, 984} & -0.201\, 343 & 0.804\, 447 &
   1.777\, 197 & 1.787\, 578 \\
11 & \ul{1.997\, 155} & \ul{2.910\, 243} & -0.201\, 108 & 0.803\, 798 &
   1.783\, 769 & 1.792\, 883 \\
12 & \ul{1.999\, 321} & \ul{2.936\, 268} & -0.200\, 930 & 0.803\, 270 &
   1.788\, 096 & 1.796\, 109 \\
13 & \ul{2.000\, 587} & \ul{2.953\, 949} & -0.200\, 792 & 0.802\, 839 &
   1.791\, 041 & 1.798\, 111 \\
14 & \ul{2.001\, 321} & \ul{2.966\, 228} & -0.200\, 682 & 0.802\, 484 &
   1.793\, 104 & 1.799\, 369 \\
15 & \ul{2.001\, 733} & \ul{2.974\, 923} & -0.200\, 594 & 0.802\, 190 &
   1.794\, 586 & 1.800\, 165 \\
16 & \ul{2.001\, 948} & \ul{2.981\, 188} & -0.200\, 522 & 0.801\, 943 &
   1.795\, 675 & 1.800\, 667 \\
17 & \ul{2.002\, 041} & \ul{2.985\, 770} & -0.200\, 462 & 0.801\, 735 &
   1.796\, 491 & 1.800\, 979 \\
\rd
18 & \ul{2.002\, 058} & \ul{2.989\, 170} & -0.200\, 412 & 0.801\, 557 &
   1.797\, 114 & 1.801\, 166 \\
\hline
\rud $\infty$ & \ul{2} & \ul{3} & -0.2 & 0.8 & 1.8 & 1.8 \\
\hline
\multicolumn{1}{|c||}{\rud $\delta$} & $<5\cdot 10^{-11}$ & $<8\cdot
10^{-10}$ &
$<2\cdot 10^{-11}$ & $<3\cdot 10^{-11}$ & $<3\cdot 10^{-10}$ &
$<3\cdot 10^{-10}$ \\
\hline
\end{tabular}
\end{center}
\end{table}
} 

Now, let us turn to the excitations.  We are going to present
numerical data of the lowest energy levels which in the finite-size
scaling limit contribute to the partition functions $\D{1}{0}\zz =
\D{0}{3/2}\zz =$ \mbox{$ \C{1}{1}(z)\,\C{1}{1}(\zquer)$}, $\D{2}{0}\zz
= \D{0}{1/2}\zz =$ \mbox{$ \C{1}{2}(z)\,\C{1}{2}(\zquer)$}, and
$\D{2}{1}\zz = \D{4}{1/2}\zz =$ \mbox{$
\C{1}{1}(z)\,\C{1}{2}(\zquer)$} (see eq.~(\ref{3e42})).  In what
follows, we will consider energy eigenvalues only since the scaled
momenta (see eqs.~(\ref{3e7}) and (\ref{3e7a})) of the levels (taking
into account possible shifts of $N/2$ corresponding to a shift of
$\pi$ in the momentum) are always equal to their infinite-size limit
(which clearly coincides for even and odd number of sites in the
corresponding sectors). This also explains why we need not consider
the sectors \mbox{$\D{3}{1}\zz = \D{1}{1/2}\zz =
\C{1}{2}(z)\,\C{1}{1}(\zquer)$} because the corresponding energy
spectra for finite chains are identical to those of the sectors
\mbox{$\D{2}{1}\zz = \D{4}{1/2}\zz = \C{1}{1}(z)\,\C{1}{2}(\zquer)$}
(the corresponding Hamiltonians are related by complex conjugation).
Tables~\ref{tab2}--\ref{tab3} show the lowest scaled energy gaps that
contribute to the partition functions (\ref{3e42}).  Underlined values
correspond to exactly degenerate eigenvalues. We also extrapolated the
finite-size values to infinite length using the algorithm of
ref.~\cite{StoBur} (see also \cite{HS}) where the free variable of the
extrapolation algorithm was chosen to be 2. The values of $\delta$
given in Table~\ref{tab2} and \ref{tab3} are the absolute difference
between the extrapolated and exact scaling dimensions.  Obviously, the
agreement between extrapolated and exact data is extremely good. This
observation is in perfect agreement with our conjecture that even in
finite systems there is no distinction necessary between an even and
an odd number of sites for this model. Hence, the half-integer charge
sectors of the XXZ Heisenberg chain do not correspond to new boundary
conditions here.

{\small
\begin{table}[tb]
\caption{\label{tab3} \protect\footnotesize
The lowest scaled energy gaps that contribute to the
sectors $\D{2}{0}$ resp. $\D{0}{1/2}$ are given
for chains of length $N$. Also shown are
corresponding exact values for $N\rightarrow\infty $
and the differences $\delta$ between the exact and the
extrapolated values (see text for details).}
\begin{center}
\begin{tabular}{|r||l|l|l|l|l|l|}
\hline
\multicolumn{1}{|c||}{\rud $N$} &
\multicolumn{6}{|c|}{$\C{1}{2}(z)\,\C{1}{2}(\zquer )$} \\
\hline
\ru
 2 & -0.449\, 035 &     0.200\, 81   &                  &             &
                    &                  \\
 3 & -0.416\, 278 & \ul{0.487\, 387} &                  &             &
                    &                  \\
 4 & -0.408\, 337 & \ul{0.555\, 038} &     1.051\, 478  & 1.766\, 633 &
                    &                  \\
 5 & -0.405\, 135 & \ul{0.577\, 995} & \ul{1.314\, 348} & 1.719\, 483 &
                    & \ul{2.552\, 737} \\
 6 & -0.403\, 498 & \ul{0.587\, 571} & \ul{1.440\, 657} & 1.686\, 862 &
       1.795\, 216  & \ul{2.628\, 080} \\
 7 & -0.402\, 541 & \ul{0.592\, 214} & \ul{1.505\, 180} & 1.665\, 366 &
   \ul{2.084\, 961} & \ul{2.649\, 283} \\
 8 & -0.401\, 932 & \ul{0.594\, 735} & \ul{1.540\, 342} & 1.650\, 769 &
   \ul{2.260\, 334} & \ul{2.652\, 827} \\
 9 & -0.401\, 519 & \ul{0.596\, 227} & \ul{1.560\, 652} & 1.640\, 489 &
   \ul{2.369\, 054} & \ul{2.650\, 289} \\
10 & -0.401\, 226 & \ul{0.597\, 172} & \ul{1.572\, 990} & 1.633\, 008 &
   \ul{2.438\, 439} & \ul{2.645\, 856} \\
11 & -0.401\, 011 & \ul{0.597\, 804} & \ul{1.580\, 817} & 1.627\, 406 &
   \ul{2.484\, 026} & \ul{2.641\, 109} \\
12 & -0.400\, 848 & \ul{0.598\, 246} & \ul{1.585\, 973} & 1.623\, 109 &
   \ul{2.514\, 805} & \ul{2.636\, 636} \\
13 & -0.400\, 721 & \ul{0.598\, 567} & \ul{1.589\, 479} & 1.619\, 743 &
   \ul{2.536\, 112} & \ul{2.632\, 626} \\
14 & -0.400\, 621 & \ul{0.598\, 806} & \ul{1.591\, 933} & 1.617\, 058 &
   \ul{2.551\, 201} & \ul{2.629\, 108} \\
15 & -0.400\, 541 & \ul{0.598\, 990} & \ul{1.593\, 693} & 1.614\, 885 &
   \ul{2.562\, 107} & \ul{2.626\, 049} \\
16 & -0.400\, 475 & \ul{0.599\, 134} & \ul{1.594\, 984} & 1.613\, 100 &
   \ul{2.570\, 138} & \ul{2.623\, 397} \\
17 & -0.400\, 420 & \ul{0.599\, 248} & \ul{1.595\, 949} & 1.611\, 617 &
   \ul{2.576\, 150} & \ul{2.621\, 097} \\
\rd
18 & -0.400\, 375 & \ul{0.599\, 341} & \ul{1.596\, 683} & 1.610\, 372 &
   \ul{2.580\, 721} & \ul{2.619\, 097} \\
\hline
\rud $\infty$ & -0.4 & \ul{0.6} & \ul{1.6} &
 1.6 & \ul{2.6} & \ul{2.6} \\
\hline
\multicolumn{1}{|c||}{\rud $\delta$} & $<3\cdot 10^{-12}$ & $<2\cdot
10^{-11}$ &
$<2\cdot 10^{-10}$ & $<3\cdot 10^{-12}$ & $<5\cdot 10^{-10}$ &
$<2\cdot 10^{-13}$ \\
\hline
\end{tabular}
\end{center}
\end{table}
} 

\section{Conclusions}
\setcounter{equation}{0}

In ref.~\cite{toro} it was shown that the finite-size scaling spectra
of the XXZ Heisenberg chain with toroidal boundary conditions and an
even number of sites contain the spectra of various series of models
with central charge less than one, all of them belonging to the
unitary series. A projection mechanism was presented that allowed for
the explicit extraction of the spectra for each model in the
finite-size scaling limit.  The idea of the projection mechanism is
first to choose an excited state of the (normalized) XXZ Heisenberg
chain (\ref{3e2}) with an anisotropy given by $q$ and some boundary
angle $\alpha_0$ as the new ground state of the projected system. The
central charge of the projected model is related to $q$ and $\alpha_0$
by eq.~(\ref{3e15}).  In a next step we fix a specific relation
between $q$ and $\alpha_0$ (\ref{3e19a}). This defines classes of
models ($R$ and $L$ models) and their physical properties. Finally, we
presented an subtraction algorithm, the projection mechanism, which
allows for the extraction of the finite-size scaling spectra of the
projected systems from the XXZ Heisenberg chain by properly choosing
charge sectors and boundary conditions and taking differences of sets
of eigenvalues in these sectors. The levels that remain after this
projection are the energy levels of the projected models.

It was observed that under certain circumstances, i.e., for certain
choices of the anisotropy parameter $q$ and the boundary twist
$\alpha$, the same mechanism also works on finite chains. This means
that all the eigenvalues of certain sectors are exactly degenerate
with part of the energy levels of the sectors from which the former
are subtracted according to the projection rules derived in the
finite-size scaling limit. Subsequently this important observation
could be traced back to properties of the quantum algebra \UQSU\
\cite{PS}.

In this paper we generalized these results in several aspects.

In the $1_R$ ($1_L$) series of ref.~\cite{toro}, the low-temperature
$O(p)$ models ($O(p)$ models) with central charge $c= 1-6/(m(m+1))$
only the cases $m$ even (odd) were considered.  Here, we completed
this work by considering all values of $m$ for both series, here
called $R=2$ ($L=2$) models. The missing models were shown to have the
same sectors as the corresponding $R=1$ (Potts) models ($L=1$
(tricritical Potts) models) with the same central charge. The exact
degeneracies observed in these models accounting for the possibility
to apply the projection mechanism in some (not all) sectors even on
finite chains were explained by the action of \UQSU\ (sec.~3). This is
an interesting generalization of the results of \cite{PS} who only
discuss what we call the $R=1$ series. That the projection mechanism
works exactly in finite chains in some sectors but not in all of them
needs further understanding.

Furthermore, we generalized the projection mechanism to arbitrary real
values of $m$, i.e., to all non-unitary minimal and non-minimal models
with $c<1$ in the $R$ and $L$ series. In this way we were able
generate all possible sectors of these models according to their
internal global symmetries and the resulting toroidal boundary
conditions.  As in the unitary series the sectors of the non-unitary
$R=1,2$ and $L=1,2$ models with the same central charge are in general
different. This implies that the physical meaning of the operators in
these systems is not specified by their anomalous dimension alone and
hence universality classes are {\em not}\/ completely defined by the
central charge and some set of anomalous dimensions.

Extending the work of ref.~\cite{toro}, we considered chains with an
odd number of sites giving rise to half-integer charge sectors. In
these sectors we discovered in some of the projected models anomalous
dimensions of so far unknown spinor fields. We studied the new sectors
of the Ising model (\ref{3e45a}) and the $3$-state Potts model
(\ref{3e45b}) in more detail (sec.~4.1).  We showed that at their
self-dual point, these two models have an additional symmetry, the
duality transformation.  This symmetry gives rise to a new class of
``duality twisted'' toroidal boundary conditions. This means that the
Hamiltonian of the corresponding models commutes with a generalized
translation operator (\ref{hatt}) performing a duality transformation
at the boundary in combination with a translation. This is in close
analogy to the well-known antiperiodic boundary conditions in the
Ising model where the generalized translation operator (\ref{TISA})
performs a spin-flip at the boundary combined with a translation. The
form of the projection mechanism suggests that most of the projected
systems have this additional symmetry. However, we also discovered
(non-unitary) systems where the spectra for integer and half-integer
charge sectors are identical (sec.~4.2). As the integer charge sectors
include the sectors with periodic boundary conditions (in the
projected system), these appear to be models where the duality twist
is identical to periodic boundary conditions and, as a consequence,
the duality transformation the identity operator.  This phenomenon is
unknown in the unitary series.

Using the representation theory of the quantum algebra \UQSU\ and
intertwining relations between the quantum algebra generators and
different sectors of the Hamiltonian of the XXZ Heisenberg chain and
the corresponding translation operator (see appendix~A), we were in
fact able to explain {\em all}\/ degeneracies observed numerically in
ref.~\cite{toro} and in addition we found similar symmetry properties
for many of the new sectors (sec.~3). In particular, for all sectors
(including the half-integer charge sectors) of all $R=1$ models the
projection procedure as described in this paper works for finite
chains and the spectrum of the finite-size projected systems therefore
is explicitly known. Furthermore, also the equality of the
corresponding momentum eigenvalues of the degenerated levels could be
proved (up to possible shifts of $\pi$ in the momentum).

U.~G.\ would like to thank M.~Baake, B.~Davies, O.~Foda, B.~M.~McCoy,
P.~A.~Pearce, V.~Rittenberg, M.~Scheunert, and Y.-K.~Zhou for valuable
discussions.  We gratefully acknowledge Research Fellowships of the
Deutsche Forschungsgemeinschaft (DFG) (U.G.) and the Minerva
foundation (G.S.).

\appendix
\renewcommand{\theequation}{\Alph{section}.\arabic{equation}}
\renewcommand{\thesection}{Appendix~\Alph{section}:}
\renewcommand{\thesubsection}{\arabic{subsection})}
\setcounter{section}{0}
\section{Coincidences in the spectra and the quantum algebra  
U$_{\mbox{{\boldmath $q$}}}$[sl(2)]}
\setcounter{equation}{0}

In what follows, we show how the observed degeneracies in the spectra
of the XXZ chain with various toroidal boundary conditions follow from
the representation theory of the quantum algebra \UQSU .  This is
achieved by establishing explicit intertwining relations between
elements of \UQSU\ and different sectors of the XXZ - Hamiltonian with
toroidal boundary conditions.  In addition, the equality of the
momenta of the degenerate levels is proved by an analogous argument.

Let us commence defining the quantum algebra \UQSU . It is generated
by the four generators $\Spm$ and $q^{\,\pm\Sz}$ subject to the
relations (see \cite{PS} and references therein)
\begin{equation} 
q^{\,\Sz} \Spm \; = \; q^{\,\pm 1}\Spm q^{\,\Sz} , \qquad
\left[ S^{+},S^{-} \right]\; =\; [2\Sz ]_{q}
\label{2e12} \end{equation}
(together with the relation $q^{\,\Sz}q^{\, -\Sz}=q^{\, -\Sz}q^{\,\Sz}=1$
which has been anticipated by the notation),
where $[x]_{q}$, the ``$q$-deformed of $x$'', is defined by
$[x]_{q}=(q^{x}-q^{-x})/(q-q^{-1})$
and $q$ is a complex number with $q^{2}\neq 1$.
A representation of this quantum algebra on
\mbox{${\cal H}(N) \cong  (\CC^{2})^{\otimes N}$}
in terms of Pauli matrices is given by \cite{PS}
\begin{equation} \begin{array}{rcl}
{\displaystyle\Spm} & = & {\displaystyle\sum_{j=1}^{N} S^{\pm}_{j}} 
        \; =\; {\displaystyle \frac{1}{2} \sum_{j=1}^{N}\;
           q^{\,\frac{1}{2} \sum_{k=1}^{j-1} \sz{k}} \, \spm{j} \,
                  q^{-\frac{1}{2} \sum_{k=j+1}^{N} \sz{k}}} \\
{\displaystyle q^{\,\Sz}}
           & = & {\displaystyle q^{\,\frac{1}{2} \sum_{j=1}^{N} \sz{j}}  .}
\end{array} \label{2e14} \end{equation}

The decomposition of this in general reducible representation on
${\cal H(N)}$ into irreducible representations of \UQSU\ has been
discussed in ref.~\cite{PS}.  It turns out that for generic values of
$q$, i.e., $q$ not being a root of unity, this decomposition resembles
the undeformed \USU\ case \cite{LU1,RO} whereas for $q$ being a root
of unity the situation is completely different \cite{LU2}.  Consider
the case $|q|\neq 1$ where one has a continuous dependence on the
complex variable $q$. If $q$ approaches a root of unity, two formerly
irreducible representations may become connected by the action of
\UQSU\ constituting one larger indecomposable representation in this
way.  These ``mixed'' representations however are no longer
irreducible since they contain the smaller of the two representations
as an irreducible submodule, i.e., the representation (\ref{2e14}) in
this case is no longer completely reducible.

Following an idea suggested in ref.~\cite{PS}, we now calculate the
action of powers of the operators $\Spm$ on the Hamiltonian
$H(q,\alpha ,N)$ (\ref{2e1}) and the translation operator $T(\alpha
,N)$ (\ref{2e5}).  The other generator $q^{\,\Sz }$ obviously commutes
with both $H(q,\alpha ,N)$ and $T(\alpha ,N)$. By induction one can
show that
\begin{eqnarray}
\frac{(S^{\, \pm})^{n}}{[n]_{q}!} H(q,\alpha ,N) & = &
H(q,q^{-2n}\alpha, N) \frac{(S^{\, \pm})^{n}}{[n]_{q}!} \nonumber \\
& & \hspace{-8ex} 
\mp 2\left( \sz{N} q^{-\sz{N}} S^{\,\pm}_{1}
\frac{(q^{\,\pm 1} S^{\,\pm})^{n-1}}{[n-1]_{q}!} - 
S^{\,\pm}_{1} S^{\,\pm}_{N}
\frac{(S^{\,\pm})^{n-2}}{[n-2]_{q}!} \right)
\left( 1 - q^{\, 2(\Sz\pm n)}\alpha^{\,\mp 1} \right) \nonumber \\
& & \hspace{-8ex}
\mp 2\left( \sz{1} q^{\,\sz{1}} S^{\,\pm}_{N}
\frac{(q^{\,\mp 1}S^{\,\pm})^{n-1}}{[n-1]_{q}!} - S^{\,\pm}_{N} S^{\,\pm}_{1}
\frac{(S^{\,\pm})^{n-2}}{[n-2]_{q}!} \right)
\left( 1 - q^{-2(\Sz\pm n)}\alpha^{\,\pm 1} \right) \nonumber \\
& & \label{2e15}  \\
\frac{(S^{\,\pm})^{n}}{[n]_{q}!} T(\alpha ,N) & = &
(q^{\sz{1}})^n T(\alpha, N) \frac{(S^{\,\pm})^{n}}{[n]_{q}!} \nonumber \\
& & + S^{\,\pm}_{1} \frac{(q^{\,\pm 1} S^{\,\pm})^{n-1}}{[n-1]_{q}!}
    T(\alpha ,N) \left( 1 - q^{\, 2(\Sz\pm n)} 
\alpha^{\,\mp 1} \right) \nonumber
\\
& = &
\varepsilon T(q^{-2n}\alpha, N) \frac{(S^{\, \pm})^{n}}{[n]_{q}!} \nonumber \\
& &  + S^{\,\pm}_{1} \frac{(q^{\,\pm 1} S^{\,\pm})^{n-1}}{[n-1]_{q}!}
    T(\alpha ,N) \left( 1 - q^{\, 2(\Sz\pm n)} \alpha^{\,\mp 1} \right)
\label{2e16}
\end{eqnarray}
with $\varepsilon =\pm 1$. Here, we used $[n]_{q}!$ for the product
\mbox{$[n]_{q}! =\prod_{j=1}^{n} [j]_{q}$}.  For the case $n=1$,
eq.~(\ref{2e15}) has already been obtained in ref.~\cite{PS}. The
ambiguity in eq.~(\ref{2e16}) is due to the square root (cf. the
definition of the translation operator in eq.~(\ref{2e5})), the sign
being fixed (but depending on the actual values of $\alpha$, $q$ and
$n$) once one chooses a particular branch.  We do not want to
investigate this further, we rather consider all momenta to be defined
modulo $\pi$ (which after all is exactly what enters in the
finite-size scaling partition function (see
eqs.~(\ref{3e5})--(\ref{3e7})).

It is our aim to extract from the above equations information about
degeneracies in the spectra of $H(q,\alpha ,N)$ for different boundary
conditions $\alpha$.  For this purpose it is convenient to consider
the several charge sectors separately. With the abbreviations
\begin{equation}
H^{K}_{Q}(N) \; = \; H(q,q^{\, 2K},N) \cdot {\cal P}_{\, Q}(N),\qquad
T^{K}_{Q}(N) \; = \; T(q^{\, 2K},N) \cdot {\cal P}_{\, Q}(N) 
\label{2e20} \end{equation}
we obtain from eqs.~(\ref{2e15}) and (\ref{2e16}) the following
``intertwining relations''
\begin{eqnarray}
\frac{(S^{\,\pm})^{n}}{[n]_{q}!} H^{Q+n}_{\pm Q}(N) & = &
H^{Q}_{\pm (Q+n)}(N) \frac{(S^{\,\pm})^{n}}{[n]_{q}!} \label{2e22}\\[1ex]
\frac{(S^{\,\pm})^{n}}{[n]_{q}!} T^{Q+n}_{\pm Q}(N) & = &
\varepsilon T^{Q}_{\pm (Q+n)}(N) \frac{(S^{\,\pm})^{n}}{[n]_{q}!} 
\label{2e23}
\end{eqnarray}
for all $n=1,2,3,\ldots $ and $-\frac{N}{2}\leq Q\leq \frac{N}{2}$,
where \mbox{$\varepsilon = \pm 1$}.

Suppose now that the set $\left\{ v_{j}^{Q}\in {\cal H}_{\, Q}(N)\;
,\; j=1,2,\ldots ,\DIM \right\}$ of common eigenvectors of
$H^{K}_{Q}(N)$ and $T^{K}_{Q}(N)$ constitutes a basis of ${\cal H}_{\,
Q}(N)$ and denote the set of pairs of corresponding eigenvalues by
$\EH{K}{Q}(N)$
\begin{equation}
\EH{K}{Q}(N)\; =\; \left\{ (e_{j},p_{j})\;\mid\;
H^{K}_{Q}(N) v^{Q}_{j} = e_{j} v^{Q}_{j} \; , \;
T^{K}_{Q}(N) v^{Q}_{j} = \pm\exp(-ip_{j}) v^{Q}_{j} \right\}  .
\label{2e26} \end{equation}
Then by applying eqs.~(\ref{2e22}) and (\ref{2e23}) to eigenstates of
the Hamiltonian and the translation operator one obtains the
inclusions
\begin{equation}
\EH{Q+n}{\pm Q}(N)  \; \supset  \;
\EH{Q}{\pm (Q+n)}(N)  \label{2e27}
\end{equation}
for all positive values of $Q$ and $n$.  Strictly speaking these
inclusions are obtained for generic values of $q$ only but since the
eigenvalues depend continuously on $q$ (in the case of finite $N$)
they also hold for the case that $q$ is a root of unity.  Indeed in
this case the observed degeneracies are much higher than for generic
$q$.

To obtain the inclusions (\ref{2e27}) we in fact also used the simple
structure of the irreducible representations of \UQSU\ for generic
$q$. One only has to realize that if one has a level in charge sector
$Q$ inside any irreducible representation, this representation
contains exactly one level in the charge sectors $Q^{\prime}$ with
$|Q^{\prime}| \leq Q$ where $Q^{\prime}$ is integer (half-integer) if
$Q$ is integer (half-integer).

Using in addition the charge conjugation transformation $C$
(\ref{2e10}) which leads to $\EH{K}{Q}(N)\equiv\EH{-K}{-Q}(N)$ for all
values of $K$ and $Q$, one finally obtains the following result:
\begin{equation}
\EH{K}{Q}(N) \;\supset\; \EH{Q}{K}(N)
 \mbox{for all }\; |K| \geq |Q|
\label{2e30} \end{equation}
and $K$ and $Q$ are both integer (half-integer) numbers for even (odd)
number of sites, respectively.

Now we turn to the case when $q$ is a root of unity.  We define an
integer $p$ by
\begin{equation}
p\; =\; \min \left\{ n>0\;\mid\; q^{2n}=1 \right\}
 \; =\; \min \left\{ n>0\;\mid\; q^{n\rule{0mm}{2mm}}\in
    \{\pm 1\}\right\}  .
\label{2e31} \end{equation}
In this case the boundary condition is defined modulo $p$ since
$H^{K}_{Q}(N) = H^{K+\nu p}_{Q}(N)$ and
$T^{K}_{Q}(N) = T^{K+\nu p}_{Q}(N)$
for all $\nu\in\ZZ$. Hence from eq.~(\ref{2e30})
one directly obtains the result
\begin{equation}
\EH{K}{Q}(N)\; \supset\; \EH{Q}{K+\nu p}(N)
\label{2e34} \end{equation}
for all $K\geq |Q|$, where $\nu$ is an arbitrary positive integer
number and we restrict $K$ to the values $0\leq K\leq p-1/2$.

In addition to this rather trivial modification one observes
additional degeneracies which show up due to the mixing of irreducible
representations of \UQSU\ when $q$ approaches a root of unity, since
all states in a indecomposable representation that correspond to
eigenstates of our Hamiltonian in the appropriate sectors have the
same energy (and momentum). To make the statement more precise, we use
the results of ref.~\cite{PS} on the indecomposable representations of
\UQSU\ for $q$ being a root of unity. This finally leads to the
following inclusion relation
\begin{equation}
\GH{K}{Q}(N) \;\supset\;
\GH{Q}{K}(N) \;\equiv\;
\GH{-Q}{-K}(N)
\label{2e35}\end{equation}
for $\frac{p}{2} \geq |K| \geq |Q| \geq 0$  and
$2 K \equiv 2 Q \equiv N \bmod 2$, where
$\GH{K}{Q}(N)$ is defined
to be
\begin{equation}
\GH{K}{Q}(N) \; =\; \bigcup_{\nu\in\ZZ}
\EH{K}{Q+n\nu}(N).
\label{2e36}\end{equation}
By closer inspection of eq.~(\ref{2e35}) one realizes
(see ref.~\cite{PS}) that the set of all levels contained in
$\GH{K}{Q}(N)$ which are not contained in
$\GH{Q}{K}(N)$,
i.e., the difference $\DDH{K}{Q}(N)$ of the two sets
\begin{equation}
\DDH{K}{Q}(N) \; = \; \GH{K}{Q}(N)\; -\;
\GH{Q}{K}(N)  ,
\label{2e37}\end{equation}
is just the set of levels that correspond to the highest weights of
those irreducible representations in the decomposition of the
representation (\ref{2e14}) which are isomorphic to \USU\ irreducible
representations. It is therefore obvious that eq.~(\ref{2e36}) really
relies on the structure of the irreducible representations and
contains more information about the spectrum than eq.~(\ref{2e34}).

\section{The duality transformation in the Ising quantum chain}
\setcounter{equation}{0}

Here we give a brief review of the duality transformation in the Ising
quantum chain. The duality transformation is a map between the low and
high temperature phases of the system. Denoting the inverse
temperature by $\lambda$ and the Hamiltonian (to be specified below)
by $H(\lambda)$, one finds that $H(\lambda)$ and its dual
$H^D(\lambda)$ have the same spectrum \cite{Kogut} and are related by
the duality transformation $D$
\begin{equation}\label{ddual}
D \; H(\lambda) \; =\;  H^D(\lambda) \; D  .
\end{equation}
$D$ depends on the boundary conditions. In what follows we consider
the mixed sector Hamiltonian $\ol{H}$ of a chain with $M$ sites
\begin{equation}\label{dhmix}\begin{array}{rcl}
\ol{H}^{\, (\pm)}(\lambda) & = &
\displaystyle - \,\left\{ \,\sum_{j=1}^{M-1}\,
\left( e_{\, 2j-1}-\frac{1}{2} \right) + \lambda
\left( e_{\, 2j}-\frac{1}{2} \right) \right.   \\
 & & \displaystyle\left. \;\; + \:
\left( e_{\, 2M-1}-\frac{1}{2} \right) + \lambda
\left( e_{\, 2M}^{\, (\pm)}-\frac{1}{2} \right)
\rule{0mm}{8mm} \right\}
\end{array}\end{equation}
where the $e_{\, j}$ for $1\leq j\leq 2M-1$ are defined by
eq.~(\ref{TLAIS2}) and $e_{\, 2M}^{\, (\pm)} = (1 \pm
S\sz{M}\sz{1})/2$.  $S$ is the spin-flip operator defined in
eq.~(\ref{Flip}). In $\ol{H}^{\, (+)}$ one has periodic boundary
conditions in the even charge sector ($S=1$) (for brevity, we denote
the projection on this sector by $H_{0}^{0}$) and antiperiodic
boundary conditions in the odd sector ($S=-1$) denoted by
$H_{1}^{1}$. $\ol{H}^{\, (-)}$ corresponds to periodic boundary
conditions in the odd sector (denoted $H_1^0$) and antiperiodic
boundary conditions in the even sector $H_0^1$.  $\ol{H}^{\,
(\pm)}(\lambda)$ satisfies the duality relation
\begin{equation}\label{dual}
D^{\, (\pm)}\;  \ol{H}^{\, (\pm)}(\lambda)\; (D^{\, (\pm)})^{-1} 
\; = \; {\ol{H}^{\, (\pm)}}^D(\lambda)  
\; = \; \lambda\, \ol{H}^{\,\pm} \left( \frac{1}{\lambda} \right)  .
\end{equation}
The duality transformation $D^{\, (\pm)}$ is given by \cite{TLA,DL}
\begin{equation}\label{dut}
D^{\, (+)} \; = \; \OP{j=1}{2M-1} g_{\, j},\qquad
D^{\, (-)} \; = \; D^{\, (+)} \sz{M}  .
\end{equation}
The operators $g_{\, j}$ are related to the $e_{\, j}$ by \mbox{$g_{\,
j} = (1 + i) e_{\, j}-1$}.  The $g_{\, j}$ are invertible (one finds
$g_{\, j}^{\, -1} = g_{\, j}^{\,\ast}$) and one can show that $D$ acts
on the $e_{\, j}$ as follows \cite{DL,GS}
\begin{equation}\label{xxx}
\begin{array}{rcl}
D^{\, (\pm)}\: e_{\, j}\: (D^{\, (\pm)})^{-1} & = & e_{\, j+1}\qquad 1
\leq j \leq 2M-2 \\ D^{\, (\pm)}\: e_{\, 2M-1}\: (D^{\, (\pm)})^{-1} &
= & e_{2M}^{\, (\pm)} \\ D^{\, (\pm)}\: e_{\, 2M}^{\, (\pm)}\: (D^{\,
(\pm)})^{-1} & = & e_{\, 1}
\end{array}\end{equation}
The $e_{\, 2j}$ are the operators dual to the $e_{\, 2j-1}$.

Using the projectors $P_{\, (\pm)}=(1\pm S)/2$ on the even and odd
sectors of $\ol{H}^{\, (\pm)}$, $P_{\, (\pm)} \ol{H}^{\, (\pm)} =
H_{l}^{l'}$ as defined above, one finds the duality relations for the
sectors of the Ising Hamiltonian
\begin{equation}
{H_{l}^{l'}}^D(\lambda) \; = \;
\lambda\, H_{l}^{l'}\left(\frac{1}{\lambda} \right) \; = \;
H_{l'}^{l}(\lambda) 
\end{equation}
with $l,l' = 0,1$.  In the derivation of these relations one has to
take into account that $D^{\, (+)}$ commutes with $P_{\, (\pm)}$, but
\mbox{$D^{\, (-)} P_{\, (\pm)} = P_{\, (\mp)} D^{\, (-)}$}. We see
that the duality relation (\ref{dual}) does not hold for the
Hamiltonian $H$ (\ref{Ising}) given in sec.~4 which in terms of the
sectors $H_l^{l'}$ is given by $H=H_0^0 + H_1^0$ for periodic boundary
conditions and $H=H_0^1 + H_1^1$ for antiperiodic boundary conditions.

From relations (\ref{xxx}) it is obvious that $D^{\, 2}$ commutes with
$\ol{H}^{\, (\pm)}$ and is related to the translation operators $T$
(\ref{TISP}) and $T'$ (\ref{TISA}).  A short calculation shows
\begin{equation}\begin{array}{ccccl}
T^{\, (+)} & = & (D^{\, (+)})^{\, 2} & = &
i^{\, M+1}\, T\, \left( P_{\, (+)}\, +\, \sx{L} P_{\, (-)} \right)  \\
T^{\, (-)} & = & (D^{\, (-)})^{\, 2} & = &
i^{\, M}\, T\, \left( P_{\, (-)}\, +\, \sx{L} P_{\, (+)} \right)  .
\end{array}\end{equation}
Since $(D^{\, (-)})^{\, 2}$ commutes with $P_{\, (\pm)}$, we can
construct a translation operator commuting with $H$ (\ref{Ising}) by
taking suitably chosen projections on the even and odd subspaces.  One
finds
\begin{eqnarray}
T & = & i^{\, -M-1} \:\left( T^{\, (+)} P_{\, (+)}\, +\,
                             i T^{\, (-)} P_{\, (-)} \right) \\
T'& = & T \sx{M}\; =\; i^{\, -M-1}\:
\left( T^{\, (+)} P_{\, (-)}\, +\, i T^{\, (-)} P_{\, (+)} \right)
\end{eqnarray}
commuting with $H$ with periodic boundary conditions and antiperiodic
boundary conditions, respectively.  The factors $i^{\, M}$, $i^{\,
M+1}$ are of course irrelevant.

Next we consider the mixed sector version of the Ising Hamiltonian
$\tilde{\ol{H}}(\lambda,\mu)$ defined by
\begin{equation}\label{dhmix2}
\tilde{\ol{H}}^{\, (\pm)}(\lambda,\mu) \; = \;
- \,\left\{ \,\sum_{j=1}^{M-1}\,
\left( e_{\, 2j-1}-\frac{1}{2} \right) + \lambda
\left( e_{\, 2j}-\frac{1}{2} \right) 
\;+\; \mu \left( e_{\, 2M-1}^{\, (\pm)}-\frac{1}{2}
\right) \right\}
\end{equation}
where the $e_{\, j}$ for \mbox{$1\leq j \leq 2M\! -\! 2$} are defined
as in eq.~(\ref{TLAIS2}) but \mbox{$e_{\, 2M-1}^{\, (\pm)} = (1 \mp
S\sy{M}\sz{1})/2$}.  The duality transformation $\tilde{D}^{\, (\pm)}$
satisfying analogous relations to (\ref{xxx}) (with $2M-1$ replaced by
$2M-2$ and $2M$ replaced by $2M-1$) is given by \cite{DL,GS}
\begin{equation}
\tilde{D}^{\, (+)} \; = \; \OP{j=1}{2M-2} g_{\, j},\qquad
\tilde{D}^{\, (-)} \; = \; \tilde{D}^{\, (+)}\: \sx{M} . 
\end{equation}
From this one obtains the transformed Hamiltonian satisfying
\begin{equation}\label{dual2}
\tilde{D}^{\, (\pm)}\: {\tilde{\ol{H}}^{\, (\pm)}}(\lambda,\lambda)\:
(\tilde{D}^{\, (\pm)})^{-1} \; = \; \lambda \tilde{\ol{H}}^{\,
(\pm)} \left( \frac{1}{\lambda},1 \right) .
\end{equation}
Since $\tilde{D}^{\, (\pm)}$ commutes with $S$, (\ref{dual2}) holds
for each sector separately. It is important to note that {\em
$(\tilde{D}^{\, (\pm)})^2$ does not commute with $ \tilde{\ol{H}}^{\,
(\pm)} (\lambda,\mu)$ unless $\lambda=\mu=1$.}  Only in this case $
\tilde{\ol{H}}^{\, (\pm)}$ is translationally invariant with
$\tilde{T}^{\, (\pm)}=(\tilde{D}^{\, (\pm)})^2$ given by
\begin{equation}\begin{array}{ccl}
\tilde{T}^{\, (+)} & = &i^{\, M}\, T\,
\left(P_{\,(+)} g_{\, 2M-2}^{\ast}g_{\, 2M-1}^{\ast}\,
-\, i P_{\, (-)} g_{\, 2M-2}g_{\, 2M-1} \right)  \\
\tilde{T}^{\, (-)} & = &i^{\, M}\, T\,
\left(P_{\, (+)} g_{\, 2M-2}g_{\, 2M-1}\,
+\, i P_{\, (-)} g_{\, 2M-2}^{\ast}g_{\, 2M-1}^{\ast} \right)
\end{array}\end{equation}
where $g_{\, 2M-2}$ and $g_{\, 2M-1}$ are defined in eq.~(\ref{gg}).
By taking projections on the even and odd subspaces one obtains the
translation operator $\tilde{T}$ (\ref{hatt}) commuting with
$\tilde{H}$ given in (\ref{HIS}):
\begin{equation}
\tilde{T}\; =\; T\, g_{\, 2M-2}\, g_{\, 2M-1}\; =\;\
 (-i)^{M} \left( T^{\, (-)} P_{\, (+)}\, +\,
                i T^{\, (+)} P_{\, (-)} \right)  .
\end{equation}
Finally, we compute $\tilde{T}^{M}$. We define \mbox{$g_{\, 0} =
-(1-i) (1 - i \sz{M} \sz{1})/2$}.  It follows from eq.~(\ref{xxx})
that $(D^{\, (+)})^{-1} g_{\, 0} D^{\, (+)} = -(1-i) (1- iS \sx{M})/2$
and therefore
\begin{eqnarray}
\tilde{T}^{\, M} & = & g_{\, 0}\,\OP{j=1}{2M-1} g_{\, j} \; = \;
            -\,\frac{1-i}{2}\, {\tilde{D}}^{\, (+)}
                  \left( 1- iS\sx{M} \right) g_{\, 2M-1} \nonumber \\
            & = & -\, {\tilde{D}}^{\, (-)} P_{\, (+)}\,
                -\, i {\tilde{D}}^{\, (+)} P_{\, (-)} .
\end{eqnarray}
This is the duality transformation in the even and odd sectors of
$\tilde{H}$.

We want to conclude this short review by noting that the duality
transformation in the $3$-state Potts model proceeds along similar
lines. Only the operators $e_{\, j}$ have to be replaced by those
given in sec.~4 for the $3$-state Potts model and the discussion of
the various sectors is slightly more complicated due to the higher
symmetry ($S_3$ instead of $S_2$). In particular, the mixed sector
version of the Hamiltonian (\ref{HP3}) is defined by taking
$\sigma_{\, M+1} = \omega^{\kappa} Z \sigma_{\, 1}$, $\kappa =0,1,2$.


\begin{thebibliography}{99}\itemsep -0.5ex
\bibitem{toro}
F.~C.~Alcaraz, U.~Grimm, and V.~Rittenberg,
\newblock  Nucl. Phys. B316 (1989) 735
\bibitem{FF}
B.~L.~Feigin and K.~B.~Fuchs,
\newblock Funct. Anal. and Appl. 16 (1982) 114
\bibitem{PS}
V.~Pasquier and H.~Saleur,
\newblock  Nucl. Phys. B330 (1990) 523
\bibitem{free1}
F.~C.~Alcaraz, M.~Baake, U.~Grimm, and V.~Rittenberg,
\newblock J. Phys. A22 (1989) L5
\bibitem{free2}
U.~Grimm and V.~Rittenberg,
\newblock Int. J. Mod. Phys. B4 (1990) 969
\bibitem{free3}
U.~Grimm and V.~Rittenberg,
\newblock  Nucl. Phys. B354 (1991) 418
\bibitem{spin1}
D.~Baranowski, G.~Sch\"{u}tz, and V.~Rittenberg,
\newblock  Nucl. Phys. B370 (1992) 551
\bibitem{DKM}
S.~Dasmahapatra, R.~Kedem, and B.~M.~McCoy,
\newblock preprint ITP-SB 92-11 (1992)
\bibitem{ABB}
F.~C.~Alcaraz, M.~N.~Barber, and M.~T.~Batchelor,
\newblock Ann. Phys. (N.Y.) 182 (1988) 280
\bibitem{oper}
F.~C.~Alcaraz, M.~Baake, U.~Grimm, and V.~Rittenberg,
\newblock  J. Phys. A21 (1988) L117
\bibitem{Y}
C. P. Yang, 
\newblock Phys. Rev. Lett. 19 (1967) 586
\bibitem{SYY}
B. Sutherland, C. N. Yang and C. P. Yang, 
\newblock Phys. Rev. Lett. 19 (1967) 588
\bibitem{McCW}
B. M. McCoy and T. T. Wu, 
\newblock Il Nuovo Cimento, Ser X, 56B (1968) 311 
\bibitem{BCR}
M.~Baake, P.~Christe, and V.~Rittenberg,
\newblock Nucl. Phys. B300 (1988) 637
 \bibitem{HB}
C.~J.~Hamer and M.~T.~Batchelor,
\newblock J. Phys. A21 (1988) L173
\bibitem{C}
J.~L.~Cardy,
\newblock {\em {\em in} Phase transitions and critical phenomena, vol.11, \\
ed. C.~Domb and J.L.~Lebowitz},
\newblock Academic Press, New York, 1987
\bibitem{RC}
A.~Rocha-Caridi,
\newblock {\em {\em in} Vertex Operators in Mathematics and Physics, \\ 
ed. J. Lepowski, S. Mandelstam and I. Singer},
\newblock Springer, New York, 1985.
\bibitem{FQS}
D.~Friedan, Z.Qiu, and S.~Shenker,
\newblock  Phys. Rev. Lett. 52 (1984) 1575
\bibitem{NRS}
B.~Nienhuis, E.~K.~Riedel, and M.~Schick,
\newblock Phys. Rev. B27 (1983) 5625
\bibitem{SALEUR}
H.~Saleur,
\newblock Phys. Rev. B35 (1987) 3657
\bibitem{GR86}
G.~v.~Gehlen and V.~Rittenberg,
\newblock J. Phys. A19 (1986) L625
\bibitem{Kogut}
J.B. Kogut,
\newblock  Rev. Mod. Phys. 51 (1979) 659
\bibitem{Chas}
P. Chaselon,
\newblock J. Phys. A22 (1989) 2495
\bibitem{TLA}
P.~Martin,
\newblock {\em Potts Models and Related Problems in Statistical
Mechanics}, World Scientific, Singapore, 1991,
and references therein
\bibitem{DL}
D. Levy,
\newblock  Phys. Rev. Lett. 67 (1991) 1971
\bibitem{GS}
G. \mbox{Sch\"{u}tz},
\newblock Weizmann preprint
\bibitem{uwe}
U.~Grimm,
\newblock Nucl. Phys. B340 (1990) 633
\bibitem{Yangs}
C.~N.~Yang and C.~P.~Yang,
\newblock Phys. Rev. 147 (1966) 303; Phys. Rev. 150 (1966) 321;
Phys. Rev. 150 (1966) 327
\bibitem{BCN}
H.~W.~J.~Bl\"{o}te, J.~L.~Cardy, and M.~P.~Nightingale,
\newblock Phys. Rev. Lett. 56 (1986) 742
\bibitem{Aff}
I.~Affleck,
\newblock Phys. Rev. Lett. 56 (1986) 746
\bibitem{StoBur}
R.~Bulirsch and J.~Stoer,
\newblock Numer. Math. 6 (1964) 413
\bibitem{HS}
M.~Henkel and G.~\mbox{Sch\"{u}tz},
\newblock J. Phys. A21 (1988) 2617
\bibitem{LU1}
G.~Lusztig,
\newblock  Adv. Math. 70 (1988) 23
\bibitem{RO}
M.~Rosso,
\newblock Commun. Math. Phys. 117 (1988) 581
\bibitem{LU2}
G.~Lusztig,
\newblock Contemporary Math. 82 (1989) 59
\end{thebibliography}
\end{document}